\documentclass [12pt]{article}
\usepackage{latexsym}
\usepackage{amsfonts}
\usepackage{amsmath}
\usepackage{acronym}
\newcommand{\be}{\begin{equation}}
\newcommand{\ee}{\end{equation}}
\newcommand{\bea}{\begin{eqnarray}}
\newcommand{\eea}{\end{eqnarray}}
\makeatletter
\@addtoreset{equation}{section}

 \makeatother
\begin{document}
\normalsize
\title{\Large "Discrete" vacuum geometry as a tool for Dirac fundamental quantization of Minkowskian Higgs model.}
\author
{{\bf L.~D.~Lantsman}\\
 Wissenschaftliche Gesellschaft bei
 J$\rm \ddot u$dische Gemeinde  zu Rostock,\\Augusten Strasse, 20,\\
 18055, Rostock, Germany; \\ 
Tel.  049-0381-799-07-24,\\
llantsman@freenet.de}
\maketitle
\begin {abstract}
We demonstrate that assuming  the "discrete" vacuum geometry in the Minkowskian Higgs model with vacuum BPS monopole solutions can justify the Dirac fundamental quantization of that model.

The important  constituting of this quantization is getting various rotary effects, including collective solid rotations inside the physical BPS monopole vacuum, and just assuming  the "discrete" vacuum geometry seems to be that thing able to justify these rotary effects.

More precisely,  assuming  the "discrete" geometry for the appropriate vacuum manifold implies the presence of thread topological defects (side by side with point hedgehog topological defects inside this manifold) in the shape of specific (rectilinear) threads: gauge fields located in the spatial region intimately near the axis $z$  of the chosen (rest) reference frame.

Just around these (topologically nontrivial) threads, collective solid rotations proceed inside the BPS monopole vacuum suffered the Dirac fundamental quantization.

The presence of topologically nontrivial threads inside the discrete vacuum manifold involves the effect annihilating equal magnetic charges, say ${\bf m}_1={\bf m}_2={\bf m}$, colliding crossing a topologically nontrivial thread.

On the other hand, in the Higgs sector of the discussed model, it is possible to construct  solutions joining in a smooth wise Higgs vacuum BPS monopole solutions. 

This submit the collective solid rotations inside the physical BPS monopole vacuum to the Gauss-shell reduction of the Minkowskian Higgs model with vacuum BPS monopoles.

It will be argued that indeed the first-order phase transition occurs in the Minkowskian Higgs model with vacuum BPS monopoles quantized by Dirac.

This comes to the coexistence of two thermodynamic phases inside the appropriate BPS monopole vacuum.

There are the thermodynamic phases of collective solid rotations and superfluid potential motions.

The important consequence of assuming the "discrete" geometry for the vacuum manifold in the Minkowskian Higgs model (with vacuum BPS monopoles) quantized by Dirac is supplanting the widespread notion about the confinement in  Yang-Mills QCD as provoked (for instance, in the maximal Abelian gauge has been fixed) by center ($\Bbb Z_2$) vortices by that taking place against   the presence therein of topologically nontrivial threads.
\end{abstract} 
 \newpage 
\tableofcontents
\newpage
\section{Introduction.}
In the recent studies \cite{fund,rem2, rem3}, the Gauss-shell reduction of the Minkowskian Higgs model with vacuum BPS monopole solutions was discussed.

Such Gauss-shell reduction is, indeed, the particular case of the  Dirac fundamental quantization \cite{Dir} of gauge theories, when such (Minkowskian) gauge model is written down in terms of topological Dirac variables \cite{David2} $A_i^D$ ($i=1,2$): transverse and gauge invariant (i.e. automatically physical) functionals of gauge fields.

As a result of this  Gauss-shell reduction, the picture of the BPS monopole vacuum as a medium possessing manifest superfluid properties and various rotary effects was given.

More exactly,  superfluid properties of the BPS monopole vacuum suffered the Dirac fundamental quantization \cite{Dir} come to its potentiality, described correctly by the {\it Bogomol'nyi} \cite{fund,rem2, rem1,LP2,LP1,Al.S.,BPS,Gold},
\be
\label{Bog}
 {\bf B} =\pm D \Phi, 
\ee
and {\it Gribov ambiguity} \cite{rem2,LP2,LP1,Pervush2},
\be
\label{Gribov.eq} [D^2 _i(\Phi _k^{(0)})]^{ab}\Phi_{(0)b} =0,
\ee
equations.

These Eqs. implicate the vacuum "magnetic" field $\bf B$ and Higgs vacuum BPS monopole modes $\Phi $ (in particular, topologically trivial modes $\Phi_{(0)a}$).

$\Phi _k^{(0)}$ are Yang-Mills (YM) vacuum BPS monopole modes, and $D$ is the (covariant) derivative.

The connection between the Bogomol'nyi and Gribov ambiguity equations is realised via the Bianchi identity $D ~B=0$ (the latter identity is equivalent to say that the vacuum "magnetic" field $\bf B$ is "transverse").

In Ref. \cite{rem1} it was pointed out to the transparent analogy between the BPS monopole vacuum and a liquid helium II specimen, possessing the manifest superfluidity.

This analogy comes, in particular, to the mathematical resemblance between the Bogomol'nyi equation (\ref{Bog}) and the expression \cite{rem1,Landau52}
\be \label{alternativ} {\bf  v}_0 =\frac{\hbar}{m} \nabla \Phi(t,{\bf r}),
\ee
for the critical velocity ${\bf v}_0$ of  superfluid potential motions inside a liquid helium II \linebreak specimen \footnote{$ v_0= {\rm min}~ (\epsilon/p)$ for the ratio of the  energy $\epsilon$ and momentum $p$ for quantum excitations spectrum in
the given liquid helium II specimen. Herewith \cite {Landau} at velocities of the liquid exceeding a {\it critical velocity} $ v_0= {\rm min}~ (\epsilon/p)$ for the ratio of the  energy $\epsilon$ and momentum $p$ for quantum excitations spectrum in
the liquid helium II, the dissipation of the liquid helium energy occurs via arising excitation quanta with momenta $\bf p $ directed antiparallel to the velocity vector $\bf v$. Such dissipation of the liquid helium energy becomes advantageous \cite{Halatnikov} just at
$$ \epsilon+ {\bf p~v}<0 \Longrightarrow \epsilon -p~v<0.$$}.

Eq. (\ref{alternativ}) implicates the phase $\Phi(t,{\bf r})$ of the complex-value helium Bose condensate wave function $\Xi (t,{\bf r})\in C$, given as \cite{rem1,Landau52}
\be
\label{Xi1}
 \Xi (t,{\bf r})= \sqrt {n_0(t,{\bf r})}~ e^{i\Phi(t,{\bf r})},
\ee

with $ n_0(t,{\bf r})$ being the number of particles in the ground energy state $\epsilon=0$.\par
$m$ is the mass of a helium atom.

On the other hand, for any potential motion, the condition \cite{rem1}
\be \label{potcon} {\rm rot}  ~{\rm grad} ~{ \Phi}=0 \ee
is always fulfilled (to within a constant) for a scalar field $\Phi$.\par
In particular, the potentiality condition (\ref{potcon}) is fulfilled for the crucial velocity ${\bf v}_0$, (\ref{alternativ}), of  superfluid potential motions inside a liquid helium II, i.e. 
$$ {\rm rot} ~{\bf v}_0=0.      $$
This implies the absence of vortices and friction forces inside the superfluid component in a liquid helium II specimen.

The similar conclusion may be drawn also about the vacuum "magnetic" field $\bf B$  in the Minkowskian Higgs model with vacuum BPS monopole solutions described correctly by the  Bogomol'nyi equation (\ref{Bog}) \cite {fund,rem2, rem1,LP2,LP1,Al.S.,BPS,Gold} (with the correction that in the latter case the operator ${\rm grad} $ would be replaced with the [covariant] derivative $D$).

\medskip
Additionally, the Gribov ambiguity equation (Gribov  equation) (\ref {Gribov.eq}) describes the Minkowskian BPS monopole
vacuum as an incompressible medium.

This fact was discussed in Ref. \cite{rem2}; it comes to the presence of typical topological invariants in the Minkowskian Higgs model with vacuum BPS monopoles quantized by Dirac \cite{Dir}.

There are the magnetic charge ${\bf m}$ and the degree of the map referring to the \linebreak $U(1)\subset SU(2)$ embedding.

\bigskip
At resolving the YM Gauss law constraint  \cite{fund,rem2, rem3,LP2,LP1,Pervush2}
\be
\label{Gauss}
\frac {\delta W}{\delta A^a_0}=0 \Longleftrightarrow [D^2(A)]^{ac}A_{0c}= D^{ac}_i(A)\partial_0 A_{c}^i
\ee 
in terms of topological Dirac variables $A^D_i$, satisfying the gauge
 \cite{David2}
\be \label{Aparallelr}
D_i^{ab} (\hat A^D) \partial_0 (\hat A_i^D)\equiv 0,
\ee
the right-hand side of  this constraint vanishes, and it becomes the homogeneous equation \cite{rem3,LP2,LP1,Pervush2}
\be \label{homo}   [D^2_i(\Phi ^{(0)})]^{ac} A_{0c}=0,        \ee 
permitting, in the Minkowskian Higgs model with vacuum BPS monopoles quantized by Dirac \cite{Dir}, the family of zero mode   solutions \cite{Pervush2, Pervush1}, 
\be \label{zero}  A_0^c(t,{\bf x})= {\dot N}(t) \Phi_{(0)}^c ({\bf x})\equiv Z^c,
   \ee
implicating the topological variable $\dot N(t)$ and Higgs (topologically trivial) vacuum Higgs BPS monopole modes $\Phi_0^a ({\bf x})$.

The topological variables $\dot N(t)$ (respectively, $N(t)$), may be specified \cite{rem3,LP2,LP1,Pervush2} via the  relation 
\bea
\label{winding num.}
\nu[A_0,\Phi^{(0)}]&=&\frac{g^2}{16\pi^2}\int\limits_{t_{\rm in} }^{t_{\rm out} }
dt  \int d^3x F^a_{\mu\nu} \widetilde{F}^{a\mu \nu}=\frac{\alpha_s}{2\pi}  \int d^3x F^a_{i0}B_i^a(\Phi^{(0)})[N(t_{\rm out}) -N(t_{\rm in})]\nonumber \\
 &&  =N(t_{\rm out}) -N(t_{\rm in})= \int\limits_{t_{\rm in} }^{t_{\rm out} } dt \dot N(t), 
 \eea 
taking account of the natural duality between the  tensors $ F^a_{i0}$ and $ F^a_{ij}$.

Herewith $\nu[A_0,\Phi^{(0)}]$ is referred to as the vacuum Chern-Simons functional, implicating the asymptotical states  "{in}"  and "{out}" taking in the time instants $t_{\rm in}$ and  $t_{\rm out}$, respectively.

As it was discussed in Ref. \cite{rem3} (repeating the arguments 
\cite{Pervush3}), it may be set
$$ t_{\rm in}\to-\infty; \quad  t_{\rm out}\to\infty. $$
The nontrivial topological dynamics inherent in the Minkowskian Higgs model with vacuum BPS monopoles quantized by Dirac comes to the specific Josephson effect \cite{Pervush3}  in the enumerated model, i.e. to the existence of  collective solid rotations inside the {\it physical} BPS monopole vacuum.

These collective solid rotations are described correctly by the free rotator action 
functional \be \label{rot} W_N=\int d^4x \frac {1}{2}(F_{0i}^c)^2 =\int dt\frac {{\dot N}^2 I}{2}\ee
involving the  rotary momentum \cite{David2}
\be \label{I} I=\int \sb {V} d^3x (D_i^{ac}(\Phi_k^0)\Phi_{0c})^2 =
\frac {4\pi^2\epsilon}{ \alpha _s}
=\frac {4\pi^2}{\alpha _s^2}\frac {1}{ V<B^2>}.    \ee
The YM coupling constant 
$$\alpha _s=\frac{g^2}{4\pi (\hbar c)^2 } $$
enters this expression for $I$ together with the typical size $\epsilon $ \cite{LP2,LP1}
 of BPS monopoles and the vacuum expectation value $<B^2>$ of the "magnetic" field.

In Ref. \cite{ rem3} there was argued that $\dot N(t)$ play the role of angular velocities of collective solid rotations inside the Minkowskian Higgs vacuum suffered the Dirac fundamental quantization \cite{Dir}.

Meanwhile, the action functional (\ref{rot}) contains vacuum "electric" monopoles (in the terminology \cite{LP2,LP1})
\be
\label{el.m}
F^a_{i0}={\dot N}(t)D ^{ac}_i(\Phi_k ^{(0)})\Phi_{(0)c}({\bf x})
\ee 
and is associated with  the real spectrum
\be \label{pin} P_N ={\dot N} I= 2\pi k +\theta; \quad \theta  \in [-\pi,\pi];  \quad k\in {\bf Z};    \ee
of the topological momentum $P_N$.

\medskip
The one of manifestations of the Josephson effect \cite{Pervush3}  occurring inside the Minkowskian Higgs vacuum suffered the Dirac fundamental quantization are the existence of  never vanishing (excluding the value $\theta=0$ of the 
$\theta$-angle) vacuum "electric" fields ("electric" monopoles) (\ref{el.m}) in the shape \cite{David2}
\be 
\label{se} F^a_{i0}\equiv E_i^a=\dot N(t) ~(D_i (\Phi_k^{(0)})~ \Phi_{(0)})^a= P_N \frac {\alpha_s}{4\pi^2\epsilon} B_i^a (\Phi _{(0)})= (2\pi k +\theta) \frac {\alpha_s}{4\pi^2\epsilon} B_i^a(\Phi_{(0)}).
       \ee 
Any vacuum "electric" field (\ref{se}) achieves its minimum in the zero topological sector of the Minkowskian Higgs model with vacuum BPS monopoles quantized by Dirac:
\be
\label{sem}   (E_i^a)_{\rm min}= \theta \frac {\alpha_s}{4\pi^2\epsilon} B_i^a; \quad -\pi\leq \theta \leq \pi.
      \ee 

\bigskip
Such is, briefly, the picture of the (topologically nontrivial) dynamics inherent in the Minkowskian Higgs model with vacuum BPS monopoles quantized by Dirac \cite{Dir}.
 
This picture seems to be correct at least at the absolute zero temperature $T=0$, when collective solid rotations inside the BPS monopole vacuum proceed in the "non-stop" regime \cite{Pervush3} and "friction forces" between this BPS monopole vacuum and its surroundings are absent.

\medskip
In Ref. \cite{rem3}, there was asserted that "geometrically", collective solid rotations inside the BPS monopole vacuum suffered the Dirac fundamental quantization  \cite{Dir} are, indeed, rotations around the infinitely narrow cylinder of the effective diameter $\sim \epsilon$ (with $\epsilon$ being the typical size of BPS monopoles) along the axis $z$  of the chosen (rest) reference frame \footnote{ For the first time, the idea of  "cylinder topologies" in (non-Abelian) gauge  theories was proposed, probably in the paper \cite{Fadd2} (where also the topological dynamical variable $N(t)$  was introduced).}. 

In the present study we attempt to ground this assertion.

This will be associated immediately with assuming the "discrete vacuum geometry" \cite{fund, rem3}
\be \label{rym1} R_{\rm YM} \equiv SU(2)/ U(1)\simeq  G_0/[U_0\otimes {\bf Z}],\ee
with
$$ \pi_1(U_0)= \pi_1(G_0)=0  $$
and 
$$ SU(2)\equiv G_0; \quad U(1)\equiv U,$$
for the degeneration space (vacuum manifold) in the Minkowskian Higgs model with vacuum BPS monopoles quantized by Dirac, involving \cite{Al.S.} the existence of (rectilinear, topologically nontrivial) threads inside the mentioned vacuum manifold, just located intimately near  the axis $z$  of the chosen (rest) reference frame.  The relation $ SU(2)\equiv G_0$ reflects the fact that the $SU(2)$ group space is simply connected (and also one-connected).

Vortices generated by such threads, similar to  those \cite{Halatnikov} one can observe in a (rested) liquid helium II specimen, may be identified naturally with collective solid rotations inside the BPS monopole vacuum suffered the Dirac fundamental quantization \cite{Dir}.

Unlike the spatial region near  the axis $z$, far off  this axis (including the region \linebreak $\vert {\bf x}\vert \to \infty$), superfluid potential motions inside the BPS monopole vacuum occur.

These motions are described correctly by the Bogomol'nyi, (\ref{Bog}), and Gribov, (\ref {Gribov.eq}), equations.

\medskip
The coexistence of (delimited somewhat in the space) collective solid rotations and superfluid potential motions inside the BPS monopole vacuum suffered the Dirac fundamental quantization is thus on hand   \cite{ rem3}.

It is the sign  of the first-order phase transition 
occurring  in the Minkowskian Higgs model with vacuum BPS monopoles quantized by Dirac. 

This phase transition is additional to the second-order one associated with the spontaneous breakdown of the initial $SU(2)$ gauge symmetry down to the $U(1)$ one.

\medskip 
In the present study we reveal the role of the discrete geometry assumed for the vacuum manifold $R_{\rm YM}$ in this first-order phase transition just as in the series of physical effects (including collective solid vacuum rotations) taking place in the Minkowskian Higgs model with vacuum BPS monopoles quantized by Dirac.

\section{Statement of problem.} 
The ensuing exposition is organized as follows. 

In {\it Section 3} (repeating partially the theses of Ref. \cite{fund}) we, utilizing the general  theory of  topological defects (see e.g. \S$\Phi$1 in \cite{Al.S.}), construct explicitly the degeneration space (vacuum manifold) $R_{YM}$ for the Minkowskian  Higgs model with vacuum BPS monopoles quantized by Dirac \cite{Dir}.

As we have seen already, it is possible to postulate the look \cite{fund}  for  this manifold, implicating the discrete multiplier ${\bf Z}$.

  Such look of the YM vacuum manifold $R_{YM}$ is 
destined by the discrete factorization \cite{fund}  of the  residual, $U(1)$, gauge symmetry group in the enumerated model:
$$ G\equiv SU(2)\simeq G_0; \quad   H \equiv  U(1) \simeq U_0 \otimes {\bf Z}.$$
\medskip
In Section 3, repeating the arguments \cite{Al.S.}, we discover three kinds of topological defects inside the vacuum manifold $R_{YM}$. There are thread and point hedgehog (stable) defects  inside this manifold. 

Herewith  thread topological defects are of a great importance for us as  those determined physical rotary effects and the first-order phase transition taking place in the Minkowskian Higgs model with vacuum BPS monopoles quantized by Dirac. We shall make sure in  this soon.

\medskip 
As it is well known \cite{Al.S.}, the sufficient condition for thread topological defects to exist in a gauge theory involving the spontaneous breakdown of the initial symmetry group
is the isomorphism
\be \label{iso} \pi_1 (R)= \pi_0 (H)\neq 0\ee
between the appropriate homotopical groups  of the degeneration space $R$ and residual symmetry group $H$ in the considered gauge theory. \par
This isomorphism is correct, in particular, for the vacuum manifold $R_{YM}$.

We ground this utilizing the holonomies group arguments (developed in Ref. \cite{LP2}; see also \cite{Al.S.}).

\medskip 
The existence of point hedgehog topological defects inside the same degeneration space $R$ is controlled by the isomorphism \cite{rem1,Al.S.}
\be \label {hedge} \pi_2 (R)= \pi_1 (H)\neq 0.\ee
These topological defects are always associated, in (Minkowskian) Higgs non-Abelian models, with various monopole solutions (typical of which will be us discussed below). Geometrically, they are located at the origin of coordinates.

\medskip 

All these suggestions should be in agreement with the assymptotical freedom of quarks and gluonic fields at distances $r\to 0$ and the confinement  behaviour of these fields at $r\sim  1~{\rm fm}$. And this matching is one of the key tasks of the Dirac fundamental quantization of the Minkowskian YM model involving vacuum BPS modes. We will often return to this task in the present study.

\bigskip The picture we drawn is very seductive and promising. However, the domain walls between different topological sectors inside the vacuum manifold can exist only at specific conditions and depending on the shape of this manifold. In turn, this shape {\it depends explicitly on the way break-down the initial gauge symmetry in the discussed model}. Now we are going to show this.

Actually, we see {\it two ways} how the initial $SU(2)$ gauge symmetry inherent into the YMH model with vacuum BPS monopole solutions quantized by Dirac is violated.

Above we have seen the supposed look (\ref{rym1}) for the vacuum manifold in this model. In this case {\it no domain walls arise}.

To show this let us consider the   long exact sequence of homotopy groups  for the fibration:

\be \label {fibr}  U_0\otimes {\bf Z} \longmapsto SU(2) \longmapsto  SU(2)/(U_0\otimes {\bf Z}).
\ee    This gives us:
\be \label {fibrh}  \ldots \longmapsto \pi_1 (U_0\otimes {\bf Z}) \longmapsto  \pi_1 (SU(2)) \longmapsto \pi_1 ( SU(2)/(U_0\otimes {\bf Z})  \longmapsto \pi_0 ( U_0\otimes {\bf Z}) \longmapsto \pi_0 (SU(2))\longmapsto \ldots
\ee  This exact sequence of homotopy groups can be simplified because of following reasons: 

$\bullet$  $\pi_1 (SU(2))=0$ ($SU(2)$ is simply connected).

$\bullet$  $\pi_0 (SU(2))=0$ ($SU(2)$ is connected).

$\bullet$ $\pi_0 (U_0\otimes {\bf Z})={\bf Z}$ (since $U_0$ is connected, $\pi_0 (U_0)=0$, and $\bf Z$ is discrete \cite{Al.S.} \footnote{The relation $\pi_0 (U_0\otimes {\bf Z})={\bf Z}$ will be us justified below.}) 

$\bullet$ $\pi_1 (U_0\otimes {\bf Z})={\bf Z}$ (since $\pi_1 (U_0)=0$ and $\pi_1 {\bf Z}={\bf Z}$ \footnote{The relation $\pi_1 {\bf Z}={\bf Z}$ follows immediately from the relation $\pi_1 S^1={\bf Z}$, widely discussed in the present study. The circle $S^1$ has its universal cover, the straight line $\cal R$. Herewith, this universal covering ${\cal R}\to S^1$ is set by the relation $p(t)=e^{2\pi i t}$. This is established by noting that every loop on the circle can be lifted uniquely to a path on the universal cover $\cal R$ and the homotopy classes of loops correspond to integers, each representing the "winding number" around the circle.}).

Thus   the simplified exact sequence is 
\be \label {simpl} 
{\bf Z} \longmapsto 0 \longmapsto  \pi_1 (SU(2)/(U_0\otimes {\bf Z}))  \longmapsto {\bf Z} \longmapsto 0
\ee
 From this sequence, we can deduce that 
\be \label{isom1} \pi_1 (SU(2)/(U_0\otimes {\bf Z})) \cong {\bf Z}.
\ee
Therefore, the fundamental group of the quotient  $G_0/(U_0\otimes {\bf Z})$ is isomorphic to  $\bf Z$. We'll follow a similar approach as before, using the long exact sequence of homotopy groups.

\medskip Let us now analyze the group $\pi_0 (SU(2)/(U_0\otimes {\bf Z}))$. We'll follow a similar approach as before, using the long exact sequence (\ref{fibrh} ) of homotopy groups.   We once again consider the fibration  (\ref{fibr}).  The relevant part of the long exact sequence for
$\pi_0$  is 
\be   
\label {lsec1}  \ldots  \longmapsto \pi_0 (U_0\otimes {\bf Z}) \longmapsto \pi_0  (SU(2) \longmapsto \pi_0 (SU(2)/(U_0\otimes {\bf Z})) \longmapsto 0.
\ee

But

$\bullet$  $\pi_0  (SU(2)=0$  ($SU(2)$ is connected)

$\bullet$   $\pi_0 (U_0\otimes {\bf Z}) ={\bf Z}$  (since $U_0$ is connected and $\bf Z$ is discrete).

Then  the simplified exact sequence is
\be   
\label {lsec11} {\bf Z}  \longmapsto 0  \longmapsto  \pi_0  (SU(2)/(U_0\otimes {\bf Z})) \longmapsto 0.
\ee

Thus from this exact sequence we see that $\pi_0  (SU(2)/(U_0\otimes {\bf Z}))$  is isomorphic to the cokernel of the map ${\bf Z}  \longmapsto 0$. But the  cokernel of ${\bf Z}  \longmapsto 0$ is precisely the  {\it trivial} group $\{ e \}$.  In other words, the vacuum manifold of us discussed model is {\it connected}, i.e.  {\it domain walls are impossible} at this its shape (\ref{rym1}). 


\bigskip In Ref. \cite{rem3}, it was also argued (implicating the general QFT arguments) that asymptotical (vacuum) states "in" and "out" (cf. (\ref{winding num.})) in the Minkowskian Higgs model with vacuum BPS monopoles quantized by Dirac would be separated by the infinite time interval \linebreak $T\to\infty $  \cite{Pervush3}. 


\bigskip
{\it Section 4}  of the present study we devote to the discussion about (topologically nontrivial) rectilinear threads containing inevitably inside the vacuum manifold $R_{YM}$.

It will be shown, repeating the arguments \cite{Al.S.}, that there are YM fields 
$$ A  _\theta(\rho, \theta, z) = A_\mu \partial x^\mu/ \partial \theta $$ 
These fields may be always represented as \cite{Al.S.}
\be 
\label{Ateta}
  A  _\theta (\rho, \theta, z)=   \exp(iM\theta) A  _\theta (\rho) \exp(-iM\theta),    \ee 
with $M$ being the generator of the group $G_1$ of global rotations compensating changes in the vacuum (Higgs-YM) configuration $(\Phi^a,A_\mu^a)$ at rotations around the  axis $z$ of the chosen (rest) reference frame. 

The elements of $G_1$ may be set as \cite{Al.S.}
\be \label {gteta} g_\theta =\exp(iM\theta).    \ee
YM fields $A_\theta $ are manifestly invariant with respect to shifts along the axis $z$.

It is important that rectilinear threads $A_\theta $ don't coincide with vacuum YM BPS monopole solutions, and, on the contrary, there are, indeed, gaps between directions of "magnetic" tension vectors: ${\bf B}_1$,
\be \label {B12} \vert {\bf B}_1 \vert \sim \partial _\rho A_\theta (\rho,\theta, z), \ee
and $\bf B$, given by the Bogomol'nyi equation (\ref{Bog}) and diverging as $r^{-2}$ at the origin of coordinates \cite{BPS}.

These gaps just testify in favour  the above discussed first-order phase transition occurring in the Minkowskian Higgs model with vacuum BPS monopoles quantized by Dirac \cite{Dir}.

\medskip
In the Higgs sector of that model, there exist \cite{Al.S.}
z-invariant (vacuum) Higgs solutions in a (small) neighborhood of the origin of coordinates:
\be \label{Higgs-teta}
  \Phi^{(n)} (\rho, \theta, z)= \exp (M\theta)~ \phi  (\rho) \quad (n\in {\bf Z}),     \ee
that can join (in a smooth wise) vacuum Higgs BPS monopoles, belonging to the same topology $n$ and disappearing \cite{David3} at the origin of coordinates. 

In the monograph \cite{Al.S.}, the claim
\be \label{clas}  \nabla_\mu \phi(\rho)\leq {\rm const}~\rho ^{-1-\delta}\quad  \delta>0;\ee
$\rho=\sqrt{x^2+y^2}$, onto the field $\phi  (\rho)$ was imposed.

Herewith, speaking "in a smooth wise", we imply that the covariant derivative $D\Phi$ of any vacuum Higgs field $\Phi_a^{(n)}$ (this derivative must  be now proportional to (\ref{clas})) merges with the covariant derivative of  a vacuum Higgs BPS monopole solution (in its topological class). 

In the present study we shall give the explicit evaluation of $\delta$. \medskip

This requirement for vacuum Higgs fields $\Phi_a^{(n)}$ to be smooth is quite natural if the goal is pursued, in the Minkowskian Higgs model with vacuum BPS monopoles quantized by Dirac \cite{Dir}, to justify various rotary effects inherent in this model.

In particular, vacuum "electric" monopoles (\ref{se}) \cite{David2} are directly proportional to \linebreak $D_i (\Phi_k^{(0)})~ \Phi_{(0)}$. 

These vacuum "electric" monopoles, in  turn, enter explicitly the action functional (\ref{rot}), describing, in the Dirac fundamental quantization scheme \cite {Dir}, collective solid rotations inside the Minkowskian BPS monopole vacuum.

\medskip
Such (smooth) sawing together  appropriate vacuum Higgs modes $\Phi^{(n)}$ and BPS monopoles is intended to remove the {\it seeming} contradiction between the manifest superfluid properties of the Minkowskian BPS monopole vacuum (suffered the Dirac fundamental quantization \cite {Dir}), setting by the Bogomolny'i, (\ref{Bog}), and Gribov ambiguity, (\ref {Gribov.eq}), equations.

Moreover, one can assert \cite {Pervush1} that
\be\label{colin} D~B\sim D~E =0  \ee
for vacuum "magnetic" and "electric" tensions in the quested Minkowskian Higgs model, i.e. that these tensions are, indeed, "transverse" vectors collinear each other.

Note that Eq. (\ref{se}) \cite{David2} just reflects this collinearity. 

Going out from this contradiction seems to be just in locating (topologically nontrivial) threads in the infinitely narrow cylinder of the effective diameter $\epsilon$ around the axis $z$ and in  joining (in a smooth wise) vacuum Higgs fields $\Phi_a^{(n)}$ and BPS monopole solutions.

In this case collective solid  rotations (vortices) inside the Minkowskian BPS monopole vacuum, occurring actually in that spatial region around the axis $z$ and described correctly by the action functional (\ref{rot}), become quite "legitimate", and simultaneously, the Gauss law constraint (\ref{homo}) is satisfied outward  this  region with smooth vacuum "electric" monopoles $E_i^a$ \cite{David2},  (\ref{se}).

The said is the one more argument in favour of the first-order phase transition occurring in the Minkowskian Higgs model with vacuum BPS monopoles quantized by Dirac and coming therein to the coexistence of two thermodynamic phases: the thermodynamic phases of collective solid  rotations and superfluid potential motions,  inside the appropriate Minkowskian BPS monopole vacuum.

Herewith the enough clear-cut picture can be observed how the enumerated thermodynamic phases are distributed inside the discrete vacuum manifold $R_{\rm YM}$.

Thread topological defects (vortices), associated with rectilinear threads $A_\theta$, are located intimately near the axis $z$  of the chosen (rest) reference frame. Actually, they refer to the cylinder of the effective diameter $d\sim 2r_1\neq 0$ with $z$ serving its symmetry axis.

Indeed, geometrically, it is the region of the vacuum manifold $R_{YM}$ possessing a nontrivial  structure determined by the value $\epsilon (0)\neq 0$.  Under these circumstances, one can suggest a highly complicated structure of vortices therein. In particular,  these have finite lengths in each topological domain.  \bigskip

Simultaneously, superfluid potential motions refer to the spatial region out of this cylinder, including the spatial region $\vert {\bf x}\vert \to\infty$ (corresponding to the infrared region of the momentum space).

\medskip
The important consequence of the presence of rectilinear threads $A_\theta $ in the Minkowskian Higgs model with vacuum BPS monopoles quantized by Dirac \cite{Dir} and involving the "discrete" vacuum geometry for the vacuum manifold $R_{\rm YM}$ is the effect 
\cite{Al.S.} of annihilating two equal magnetic charges ${\bf m}_1= {\bf m}_2={\bf m}(n)\neq 0$ ($n\in{\bf Z}$) colliding at crossing a rectilinear topologically nontrivial thread $A_\theta (n)$.



The said can lead to the  annihilation of all the topologically nontrivial YM vacuum BPS monopole modes  and excitations over  this BPS monopole vacuum (suffered the Dirac fundamental quantization \cite{Dir}) during a definite time (mathematically this can be written as $<{\bf m}>=0$).

As a consequence of such possible annihilation, Higgs (BPS monopole) modes become free electric fields: their electric charges $e$
 are dual (due to the Dirac quantization \cite{Dirac} of  electric and magnetic charges) to zero  magnetic charges can only  survive upon the above described annihilation.

Such situation when Higgs modes possess arbitrary electric charges $e$, while magnetic charges ${\bf m}\neq 0$ are confined is referred to as  the {\it Higgs phase} in modern physical literature (see e.g. \cite{Hooft}).

If quarks are incorporated in the Minkowskian Higgs model with vacuum BPS monopoles quantized by Dirac \cite{Dir},  disappearing topologically nontrivial YM modes via the "colliding" mechanism \cite{Al.S.} can cause the possibility to observe free "coloured" quarks in the spatial region near the axis $z$.

The said can serve as a (perhaps, enough rough) representation for the asymptotical freedom of quarks  in that model.


Such confinement may be provided, for instance, by surviving, upon colliding \cite{Al.S.} at threads $A_\theta$, only topologically trivial YM modes. 

\bigskip
In {\it Section 5} we discuss important consequences for QCD (based on the Minkowskian Higgs model with vacuum BPS monopoles quantized by Dirac) of assuming the "discrete" geometry.

\medskip
The thing is that the Bogomol'nyi equation (\ref{Bog}) is fulfilled to within the sign before the (covariant) derivative $D$ \cite{Gold}.

Herewith it may be shown (repeating the arguments \cite{Al.S.}), and it will be done in Section 5, that this duality in specifying the sign of the vacuum "magnetic" field $\bf B$ becomes unavoidable when there exists the (already mentioned) group $G_1$ of global rotations, compensating changes in the vacuum (YM-Higgs) configuration $(\Phi^a,A_\mu^a)$ at rotations around the  axis $z$ of the chosen (rest) reference frame.

This causes inverting the sign before the generator \cite{Al.S.} $h\equiv h(\Phi)\equiv \Phi/a$ (with $\Phi$ being a Higgs BPS monopole mode and $a^2\equiv m/\sqrt \lambda$) of the residual $U(1)$ gauge symmetry group in the quested Minkowskian Higgs model. \par
More exactly, to within an isomorphism, the elements \cite{Al.S.}
\be\label {holonomija} \alpha(n)= (P\exp(-\int \limits _0^{2\pi}A  _\theta d \theta ))^{-1}
 \ee
(with $P$ standing for parallel transports along appropriate curves [integration ways]) of the holonomies group \cite{LP2,Al.S.} $H\simeq U(1)$ may be constructed involving (topologically nontrivial) threads $ A  _\theta$.

In this case there exist gauge transformations \cite{Al.S.}
\be \label{dwaz}   \tilde h(\Phi) =\alpha(2k) h(\Phi) \alpha(2k)^{-1}=- h(\Phi)   \ee 
mapping the topological domain with a topological number $k$ in the discrete group space $H$ (and by that the appropriate topological domain inside the degeneration space $R_{\rm YM}$) into the topological domain with the topological number $-k$.\par
Herewith the change in the sign of $\bf B$ in the Bogomol'nyi equation (\ref{Bog}) occurs automatically at such mapping, accompanied also by changing the signs of magnetic charges $\bf m$. \par
Thus vacuum "magnetic" fields $\bf B$ and $-\bf B$ may be identified by means of gauge transformations (\ref{dwaz}),  and this implies, regarding the Minkowskian Higgs model with vacuum BPS monopoles quantized by Dirac, the manifest invariance of the appropriate vacuum "magnetic" energy, squared by $\vert \bf B \vert$, with respect to changes in the sign of  the vacuum "magnetic" field $\bf B$. \par
The same gauge transformations (\ref {dwaz}) influence also the "electric" energy of the Minkowskian BPS monopole vacuum (suffered the Dirac fundamental quantization \cite{Dir}), squared by the "topological derivative" $\dot N(t)$ and given by the action functional (\ref {rot}).\par
In this context, following \cite {rem3, Pervush3}, it will be useful to recall that $\dot N(t)$ can be represented as \be \label{Poi} {\dot N}(t)={\rm const} = (n_{\rm out}-n_{\rm in})/T\equiv \nu/T,        \ee 
where $n_{\rm out}$, $n_{\rm in}$~ $\in {\bf Z}$ refer to the fixed time instants $t=\pm T/2$, respectively; $\nu$ is referred to as the {\it Pontryagin index } \cite{Pervush2, Pervush3}.
\par   We see thus that the $k\to -k$ ($k\in {\bf Z}$) replacement does not change $\dot N(t)$ and therefore also the "electric" energy of the Minkowskian BPS monopole vacuum!  Note also that besides  mentioned above reversing the sign due to the gauge transformations (\ref{dwaz}), here there are natural ${\Bbb Z}_2$ discrete symmetry of the "electric" part of the action functional. In this case it is quite advisable to consider the additional ${\Bbb Z}_2$ discrete symmetry  (to the residual gauge symmetry $U_0\otimes \bf Z$, us discussed above), {\it common for the "magnetic" as well as the "electric" parts of the action functional for the vacuum manifold}.

As a result, the "electric" energy of this vacuum  and the action functional (\ref {rot}) corresponding to this "electric" energy prove to be also invariant with respect to changes in signs of topological charges $k$.\par
Just this reasoning  about the vacuum "magnetic" and "electric" energies of the \linebreak Minkowskian BPS monopole vacuum allows to draw the conclusion that the look of the initial and residual gauge symmetries groups in the Minkowskian Higgs model with vacuum BPS monopoles quantized by Dirac would be modified: respectively,
$$  U_1=U_0\otimes {\bf Z}\otimes{\Bbb Z}_2.
     $$
This implies the modification 
$$ R'_{YM}= G_0/[U_0\otimes ({\bf Z}\otimes {\Bbb Z}_2)]
      $$
in the look of the appropriate degeneration space (vacuum manifold) $R_{\rm YM}$.  It is important to note at the same time the natural isomorphism 
$$ {\bf Z}\otimes {\Bbb Z}_2 \simeq {\bf Z},$$ 
since ${\Bbb Z}_2$  is an authomorphism of the set ${\bf Z}$  of integers. 

\medskip
On the other hand, the manifest account of the $ {\Bbb Z}_2$ symmetry in the "modified" look $ U_1$ of the  residual gauge symmetry group in the Minkowskian Higgs model with vacuum BPS monopoles quantized by Dirac (inducing the look $R'_{YM}$ for the vacuum manifold in that model) via the factor ${\Bbb Z}_2$ does not involves  additional singular gauge fields.  \par 
\bigskip
Such singular gauge fields arise, for instance, in the case \cite{Lenz} of the { continuous}, by neglecting {\it centre vortices}, $SU(2)$ group geometry. \par 

The latter kind of topological defects plays an important role in the modern sight about QCD (stated, in particular, in the review \cite{Lenz} and in the papers \cite{Hugo, Maxim});  that is why the enough much place will be given centre vortices in the present study. \par
We pursue herewith the goal to demonstrate  the uselessness of centre vortices in the Minkowskian Higgs model with vacuum BPS monopoles quantized by Dirac \cite{Dir} and involving the "discrete" vacuum geometry.\par
\medskip 
Center vortices in a non-Abelian model involving, for simplicity, initial $SU(2)$ gauge symmetry group \footnote{In the present study, also the case of ${\Bbb Z}_3$ vortices, associated with the (initial) $SU(3)_{\rm col}$  gauge symmetry group in "realistic" QCD will be discussed.} at the "continuous" group geometry being assumed, are associated \cite{Lenz} with the nontrivial centre of $SU(2)$ consisting of two elements, $e$  and $-e$, with $e$ being the unit element of $SU(2)$. \par 
Herewith when one requires the centre symmetry to be present upon gauge
fixing, the isotropy group formed by the centre reflections must survive the "symmetry breakdown" induced
by the elimination of redundant variables. \par
In
this way, one can change effectively the gauge group:
$$  SU(2) \to SU(2)/\mathbb{Z}_2.     $$
So long as
$$\pi_1\big(SU(2)/\mathbb{Z}_2\big) \simeq \pi_1\big(SO(3)) = \mathbb{Z}_2,$$ 
the group space of $ SU(2)/\mathbb{Z}_2$ proves to be containing the specific kind of topological defects, referring to as center vortices. \par 
On the other hand, the real projective space 
$${\bf RP}^2\simeq S^2/\mathbb{Z}_2 $$ 
possesses the same topology that
$$ SU(2)/\mathbb{Z}_2\simeq SO(3)$$
More precisely, 
$$\pi_1\big(SU(2)/\mathbb{Z}_2\big)  =\pi_{1} ({\bf RP}^2)= {\Bbb Z}_2.$$
Furthermore, so long as ${\bf RP}^2\subset {\bf R}^3$, this implies that the group space  of $ SU(2)/\mathbb{Z}_2$ contains nontrivial singularity lines in ${\bf R}^3$ similar to those one can discover in liquid nematic crystals possessing a one symmetry axis \cite{Al.S.}. \par
\medskip
This similarity seems to be very didactic for understanding $\mathbb{Z}_2$ vortices in the YM model; that is why it will be the one of topics in Section 5. \par
In particular, it will be demonstrated (repeating the arguments \cite{Al.S.}) that the degeneration space (vacuum manifold) in a liquid nematic crystal with a one symmetry axis is indeed
  $$ R_n= SO(3)/ O(2)\simeq S^2;$$
it is the surface over which  the free energy of the crystal achieves its minimum. \par
This degeneration space corresponds to the (global) $ SO(3)$ symmetry of such a liquid nematic crystal that is violated thereupon down to its $O(2)$ subgroup. \par
On the other hand, 
$$ R_n\simeq S^2/\mathbb{Z}_2 \simeq {\bf RP}^2.$$
Just this induce topologically nontrivial singularity lines  along which the thermodynamic equilibrium is violated in the liquid nematic degeneration space $ R_n\simeq {\bf RP}^2$. They refer to as {\it disclinations} in the modern physical literature  \cite{Al.S.,Lenz}.\par
Herewith disclinations are separated by domain walls (the one in the given liquid nematic crystal) from topologically trivial lines, that may be contracted into a point; thus the appropriate degeneration space $ R_n$ is two-connected. \par 
Disclinations in liquid nematic crystals possessing a one symmetry axis are patterns of  topological defects belonging to the type of center  vortices. \par
The sign of disclinations existing in a liquid nematic crystal is the 
relation \cite{Al.S.,Lenz}
$$ \pi_{1}({\bf RP}^2)=\mathbb{Z}_{2},  $$
which one encounters also in the case of the $ SU(2)/\mathbb{Z}_2$   gauge symmetry group in the YM theory. \par
 Just the latter topological  relation allows to drawn the conclusion about the similarity of geometries in the case of liquid nematic crystals possessing a one symmetry axis and the YM theory (if it implicating, at  appropriate gauge 
fixing \cite{Lenz}, the [initial] $ SU(2)/\mathbb{Z}_2$   gauge symmetry group).

\medskip
Following \cite{Lenz}, it will be shown that singularity lines in the YM model involving the initial $SU(2)$ gauge group are always associated with purely gauge transformations 
$$ A ^{\mu}_{\mathbb{Z}_2}(x) = \frac{1}{ig}\,U_{\mathbb{Z}_2}(x)\,\partial ^{\mu}\,U_{\mathbb{Z}_2}^{\dagger}(x). $$
Herewith the gauge matrices $ U_{\mathbb{Z}_2}$ have the look \cite{Lenz}
$$  U_{\mathbb{Z}_2}(\varphi) = \exp{i\, \frac{\varphi}{2}\,\tau^3}      $$
in the cylindrical coordinates $\rho,\varphi,z,t$.\par
The gauge matrices $ U_{\mathbb{Z}_2}$ prove to be singular on the sheet $\rho = 0$ (for all $z,t$), and thus these gauge matrices exhibit the essential properties of singular gauge transformations referring to center vortices. \par
It is correctly because 
$$U_{\mathbb{Z}_2}(2\pi) = -U_{\mathbb{Z}_2}(0),$$
i.e. since any $ U_{\mathbb{Z}_2}$ gauge transformation is continuous in $SU(2)/\mathbb{Z}_2$ but discontinuous as an element of $SU(2)$. \par
The next evidence in favour of singularity of the gauge transformations
$ U_{\mathbb{Z}_2}$ on the sheet $\rho = 0$ is \cite{Lenz} the definite behaviour of Wilson loops $ W_{{\cal C}, \,\mathbb{Z}_2}$:
\be \label{Wilp} W_{{\cal C}, \,\mathbb{Z}_2} = \frac{1}{2}\,\mbox{tr}\, \big\{U_{\mathbb{Z}_2}(2\pi)\, U_{\mathbb{Z}_2}^{\dagger}(0)\big\}.    \ee
There was demonstrated in \cite{Lenz} that this expression equal to -1 for an arbitrary
path ${\cal C}$ enclosing a center vortex. \par
This Eq. allows to calculate the components of gauge fields corresponding to matrices $ U_{\mathbb{Z}_2}$ and Wilson loops $ W_{{\cal C}, \,\mathbb{Z}_2}$.\par 
These turn out to be \cite{Lenz}
$$  A^{\varphi}_{\mathbb{Z}_2}(x) = - \frac{1}{2g\rho}\tau^3, $$
manifestly singular on the sheet $\rho = 0$.
\par
Herewith singular YM fields $A^{\varphi}_{\mathbb{Z}_2}(x)$ represent correctly center vortices in the gauge model involving the $SU(2)/\mathbb{Z}_2$ symmetry group. \par 
\bigskip
In the Minkowskian Higgs model with vacuum BPS monopoles quantized by Dirac \cite{Dir} and involving the "discrete" vacuum geometry, the factor $\mathbb{Z}_2$ is also present in the explicit expression for $R'_{\rm YM}$. \par 
But the situation now is rather different than in the YM theory involving the $SU(2)/\mathbb{Z}_2$ gauge symmetry group and center vortices \cite{Lenz}. \par
In the former case, gauge transformations (\ref {dwaz}),  represented by holonomies elements $\alpha(n)$ \cite{Al.S.}, "overlap" completely gauge transformations $U_{\mathbb{Z}_2}$ \cite{Lenz} regarding changes  in signs of topological charges in the Minkowskian Higgs model with vacuum BPS monopoles quantized by Dirac. \par 
However the holonomies elements $\alpha(n)$ are associated there actually with (topologically nontrivial) thread solutions $A_\theta $ \cite{Al.S.}, i.e. with thread topological defects. \par
Just this kind of topological defects  replaces centre vortices in the mentioned \linebreak Minkowskian Higgs model, leaving no room for the latter ones. 

\bigskip
In spite of this, the discussion about center vortices seems to be very instructive: for instance, in order to understand that the Minkowskian Higgs model with vacuum BPS monopoles quantized by Dirac \cite{Dir} gives the rather new confinement picture in QCD in comparison with the one getting at assuming (as it was done, for instance, in Ref. \cite{Lenz}) the initial $SU(n)_{\rm col}/{\mathbb{Z}}_n$ ($n=2,3$) gauge group. \par 
Indeed, it may be supposed additionally that this initial $SU(n)_{\rm col} /{\mathbb{Z}}_n$ ($n=2,3$) gauge symmetry group is violated in the 
\be \label{way} SU(n)_{\rm col} /{\mathbb{Z}}_n \to U(n-1) /{\mathbb{Z}}_n \ee 
wise \footnote{In particular, 
$$ SU(3)_{\rm col}/{\mathbb{Z}}_3 \to  U(1)\otimes U(1) /{\mathbb{Z}}_3\simeq U(2) /{\mathbb{Z}}_3. $$}.\par
But irrespective of either the initial $SU(n)_{\rm col} /{\mathbb{Z}}_n$ ($n=2,3$) gauge symmetry group is exact or it is violated down to its Abelian subgroup $ U(n-1) /{\mathbb{Z}}_n $, the presence of center vortices implies \cite{Lenz,Hugo} the correct quark confinement description. \par
More precisely, it may be shown \cite{Lenz,Hugo} that the area law, serving the confinement criterion in the QCD model involving the Mandelstam linearly increasing potential \linebreak $E=K~ r$ between some quark $q$ and antiquark $\bar q$, is satisfied for Wilson loops $ U_{{\cal C}, {\mathbb{Z}_2}}$, (\ref{Wilp}), in the case of the initial $SU(2)_{\rm col}/{\mathbb{Z}}_2$ gauge symmetry group:
$$ \langle W\rangle = \sum_{n=1}^{N} (-1)^n  p_n \to \exp \big( -2\nu {\cal A}_W\big). $$
Such area law is got for the random distribution of intersection points of vortices with the enough large area $\cal A$ with those with a much smaller area ${\cal A }_W$.\par 
This random distribution has the look
\cite{Lenz, Hugo}
$$p_n = \binom{N}{n}\Big(\frac{{\cal A}_W}{{\cal A}}\Big)^n\Big(1-\frac{{\cal A}_W}{{\cal A}}\Big)^{N-n}\, .$$
Herewith $\nu$ is the density of intersection points. \par
\medskip
The  above area law remains invariable at the spontaneous breakdown, in the (\ref{way}) wise, of the $SU(2)_{\rm col}/{\mathbb{Z}}_2$ gauge symmetry. \par
In the latter case,  Wilson loops $ U_{{\cal C}, {\mathbb{Z}_2}}$, (\ref{Wilp}), contain gauge matrices $ U_{\mathbb{Z}_2}$ belonging to $ U(1) /{\mathbb{Z}}_2$ (the Abelian subgroup of $SU(2)_{\rm col}/{\mathbb{Z}}_2$). \par
Herewith the residual $ U(1) /{\mathbb{Z}}_2$ (and, generally, $ U(n-1) /{\mathbb{Z}}_n$) gauge symmetry group may be chosen by fixing the so-called {\it maximal Abelian gauge } (MAG) for YM (gluonic) fields. \par
This gauge comes to the maximal diaganalization of the appropriate gauge matrices. \par 
This may be achieved (in the simpler case of the YM theory) by means of maximizing the integral
$$ \int d^4x \left [(A_\mu^1)^2+  (A_\mu^2)^2\right ]; \quad A_\mu = A_\mu^a \tau_a. $$
It is equivalent to the fact  (see e.g. Ref. \cite{Jeff}) of maximization of the quantity
$$ \sum_x \sum_{\mu=1}^4 \mbox{Tr}[\tau_3 U_\mu(x) \tau_3 U^\dagger_\mu(x)],
   $$
where $ U_\mu(x) $ are $SU(2)$ gauge matrices. \par 
MAG may be got \cite{Sobr1} by imposing the "Lorentz" gauge
$$  D_\mu^{ab} A_\mu^b  =0.    $$
\medskip
Indeed, the combination of MAG and the center vortices in the case when the initial $SU(2)_{\rm col}/{\mathbb{Z}}_2$ gauge symmetry group is violated in the (\ref{way}) wise, is a "mix". 
\section{How one can introduce  "discrete"  geometry for degeneration spaces in  Minkowskian Higgs  models.}
Recall again the recent paper \cite{rem1}. In  this paper some modern Minkowskian Higgs  models with {\it stationary} vacuum monopole solutions were analysed. \par 
Besides already mentioned  BPS monopole model \cite{ LP2,LP1,Al.S.,BPS,Gold}, one can mention the 't Hooft-Polyakov monopole model \cite{ H-mon, Polyakov}. \par 
As it was discussed in \cite{rem1}, such (vacuum) monopole solutions are compatible with the "continuous" vacuum geometry
\be \label{contin} R\equiv SU(2)/U(1)\simeq S^2.     \ee 
Also it was argued that Minkowskian Higgs  models with { stationary} vacuum monopole solutions, involving the "continuous" vacuum geometry  (\ref{contin}), can be described quite correctly in the framework of the Faddeev-Popov (FP) "heuristic" quantization scheme \cite{ FP1}, coming \cite{ fund, rem1} to actual fixing the Weyl gauge $A_0=0$ (for instance via the multiplier $\delta(A_0)$ in appropriate FP path integrals).

\medskip
From the topological viewpoint, the {\it continuous} vacuum manifold $R$ \footnote{Following \cite{Al.S.}, $R$ may be called the {\it degeneration space}.\par

It is quite correct herewith to interpret  the initial symmetry group $G$ in a Minkowskian Higgs  model as that does not change the energy functional (Hamiltonian) of that model, while the residual symmetry group $H\subset G$ as that consisting of transformations  that keep invariant a fixed equilibrium state. \par 
All these states (at a fixed temperature $T<T_c$, with $T_c$ being the appropriate Curie point in which the initial symmetry $G$ is violated and the second-order phase transition occurs) just form the appropriate degeneration space $R=G/H$. The natural claim to this space is herewith  \it to be topological\rm. \par 
The structure of a degeneration space may be investigated with the aid of the
Landau  theory  of second-order phase transitions. \par
An equilibrium state is determined by the condition for the  
free energy of the given system  to be minimal. \par
In the
Landau  theory  of second-order phase transitions (the pattern of which is the  Minkowskian Higgs model) one supposes that an equilibrium state may be found at minimizing  the free energy of the given system  by the set of states   specified by a finite number of parameters (called \it  order parameters\rm), but not by the set of all the states. },
 (\ref{contin}), possesses the one kind of topological defects, namely the point hedgehog topological defects. \par
These topological defects are associated with vacuum monopole solutions in appropriate Minkowskian Higgs models. \par
Point hedgehog topological defects inside $R$ are determined by the topological chain \cite{rem1,LP2} 
\be
\label{point1}
\pi_2 (R)\equiv \pi_2 S^2 = \pi_ 1(H)\equiv \pi_ 1 U(1)=\bf Z. \ee
From the thermodynamic points of view, all the kinds of topological defects may be explained (repeating the arguments \cite{Al.S.}) by violating  the thermodynamic  equilibrium over a region in the appropriate coordinate (in particular, Minkowski) space. \par 
For example, point hedgehog topological defects inside $R$ are associated with violating  the thermodynamic  equilibrium in an infinitesimal neighbourhood of the origin of coordinates. Such neighborhoods are topologically equivalent to the two-sphere $S^2$. \par 

\bigskip
In the present study we attempt to demonstrate our readers that going out from the FP "heuristic" quantization scheme \cite{FP1} to the Dirac "fundamental" one \cite{Dir} in the Minkowskian Higgs model with vacuum BPS monopole solutions claims the in principle new geometrical approach to constructing the appropriate vacuum manifold (in comparison with assuming the "continuous" $\sim S^2$ vacuum geometry in the former case). \par 
This "new geometrical approach" (whose outlines were contemplated in the recent paper \cite{fund}) comes to assuming the "discrete" geometry for the vacuum manifold $SU(2)/U(1)$.  \par 
As we shall make sure soon,  this assuming explains enough good various vacuum rotary effects \cite{rem3} inherent in the physical BPS monopole vacuum suffered the Dirac fundamental quantization \cite{Dir}.\par
\medskip
So, let us represent  \cite{fund} the the residual, $U(1)\equiv H$, gauge symmetry group in the Minkowskian Higgs model with vacuum BPS monopoles quantized by Dirac in the shape of discrete space \footnote{Such representation for gauge groups spaces was proposed for the first time in the paper \cite{Pervush1}.}
\be
 \label{fact2}
    U(1) \simeq U_0 \otimes {\bf Z}.
 \ee  Such "discrete" representation for the gauge group $U(1)$ is quite justified. Really, an element of $U(1)$ can be represented as $\exp(in\theta)=\exp(i0\theta)\exp(in\theta)$ [$\exp(i0\theta)=1$]. This gives sketch to the proof  of Eq. (\ref{fact2}); however, the complete proof should use the "loop framework" from the cited Lecture 3 by \cite{Postn4}.  More precisely, if $U_0$ consists of trivial loops, i.e. those loops $u$ which are homotopical to the constant loop $e_{p_0}$  ($p_0$ is the center of the circle $S^1$) \footnote{Formally, it is the map $e: [0, 1] \rightarrow X$ such that $e(t) = x_0$ for all $t \in [0, 1]$.}, then any loop homotopical class comes to the 
\be \label{cll} [u]^n =[e_{p_0}]\circ\tau^n,\ee 
with $\tau$ being the homotopical class of loops which just once uniformly running around the circle $S^1$ and herewith counterclockwise;  $n\in {\bf Z}$. Since $[e_{p_0}]$ is the neutral element in $\pi_1 S^1$, Eq. (\ref{cll}) can be simplified, namely 

\be \label{clla}[u]^n\cong [\tau]^n.
\ee

\medskip Sometimes, when it will be convenient for us (for instance, at analysis of the topological model for the liquid helium,- a byproduct of our present discussion), we will denote the discrete manifold $ U_0 \otimes {\bf Z}$ as $\tilde U$. It is naturally ask now the question "how the isomorphism $U(1)\simeq \tilde U$ works"?  We suppose that topological sectors of the $U(1)$ group space (i.e. actually the homotopical classes $[u]^n$ of loops, (\ref{cll})) become separated, during this process, by domain walls. Herewith it can happen that such a quite separate set \cite{Engel} (defined above) is the mixture of different topologies. In this case, due to the abelian nature of $U(1)$, the topology of the domain is determined by the large in modulus topological number when all $n_i$ of the same sign.  If such topological numbers have different signs,  we merely operate with these as with integers. Thus one can assert that each separated domain possesses {\it the fixed topology}. 

Also, it can be given the alternative definition of $ \tilde U$ as {\it the disjoint union of the topological sectors} $U_1^n$  ($n\in {\bf Z}$), i.e. 
\be \label {disj}\tilde U=\bigcup _{n\in {\bf Z}} U_1^n.
\ee Now we give the sketch of the proof that  $U(1)\to \tilde U$ is indeed {\it an isomorphism} \footnote{This should provide that $\tilde U$ is manifold, {\it locally} equivalent to the topologically trivial Euclidean space $R^n$ ($n$ any natural number) but {\it globally} possessing a nontrivial topology. }. The process of gluing the sectors $U_1^n$  together  involves ensuring that the transition between adjacent sectors is smooth and continuous. This means that the boundary of each sector $U_1^n$ match up correctly with the boundary of the sectors $U_1^{n+1}$ and $U_1^{n}$.  During this gluing process, we need to check for any residual sets or gaps that might arise. These residual sets would indicate discontinuities or mismatches in the reconstruction of $U(1)$.  

To ensure that there are no residual sets, we would typically:

$\bullet$ Define a continuous map: construct a continuous map that takes each point in the disjoint union and maps it to a corresponding point on $U(1) \simeq S^1$. 

$\bullet$ Verify smooth transitions: ensure that the transition between  $U^1_n$ and $U^1_{n+1}$  is smooth, meaning there are no jumps or gaps.

$\bullet$ Topological consistency: check that the overall topology of the reconstructed space matches that of  $U(1)$.

How to construct a  smooth transition between $U_1^{n+1}$ and $U_1^{n}$? Following (\ref{cll}), the loops from the sector $U_1^{n+1}$ can be gotten from those from the  sector  $U_1^{n}$ in the shape $$[u]^{n+1} =[e_{p0}]\circ \tau ^n \circ \tau =[u]^n \circ \tau.$$
The map $\phi: \gamma_n \to \gamma_{n+1}$  (for the loops $\gamma_n \in [u]^n$ and $\gamma_{n+1} \in [u]^{n+1}$) is really an isomorphism: it is indeed  bijective mapping. To make sure of this  we recommend our readers Lecture 3 in the monograph \cite{Postn4}. Another very important for us conclusion is that the fundamental group of one-dimensional loops $\pi_1 (U(1), p_0)\simeq \tau^n$ (following\cite{Postn4}, $p_0$ is a pre-selected point on the sphere $S^1\simeq U(1)$) acts as an automorphism within the disjoint union $\tilde U$. In fact, this is equivalent to say that $\tilde U$is invariant relatively the (gauge) group $U(1)$, i.e, {\it can be treated as the degeneration space} in definite contexts, for instance in the topological $\rm He ^4$ model, discussed additionally  in the present study. 

Herewith any  homotopical class $\tau$ (possessing the topological number $n=1$ and treated \cite{Postn4} as the generator of the cyclical group  $\pi_1 (U(1), p_0)$) can be treated naturally as the operator {\it raising} the topology: from $n$  to $n+1$. In particular, such operator transforms the topological sector $U_1^n$  into  $U_1^{n+1}$. Inversely,  $\tau^{-1}$  can be treated as operator {\it 
lowering} topology: from $n$  to $n-1$.

  \bigskip    As regards the initial gauge symmetry $SU(2)$, it is one-connected and simple connected group: $$ \pi_0(SU(2))=\pi_1(SU(2))=0$$  in the framework of the natural diffeomorphism $SU(2)\simeq S^3$.  That is why, as it was announced in the previous section, we denote the initial gauge symmetry group $SU(2)$ as $G_0$.

From the topological viewpoint, the  discrete representation (\ref{fact2}) for the gauge group $H$ extracts "small" (topologically trivial) and "large" (corresponding to topological numbers $n\neq 0$) gauge transformations in the complete set of appropriate gauge transformations (the idea of such subdividing for gauge transformations was suggested in Ref. \cite {Fadd2}).\par
According to the terminology \cite {Fadd2}, the complete group $H_0$ just contains "small" gauge transformations,  that implies
\be
 \label{small}
 \pi_n H_0 =0
\ee
for loops in the group space  $H_0$ in all the dimensions $n\ge 1$.\par 
Simultaneously, in definition,
\be
 \label{oneconnect} 
 \pi_0 H_0 =0,
\ee
i.e.  $H_0$ is  the \it maximal connected component \rm \cite {Al.S.} in the gauge group $H$.\par
Later Eq. implies \cite {Al.S.} that
\be\label{dl}  \pi_0 [H_0 \otimes {\bf Z}]=\pi_0 ({\bf Z})= {\bf Z}.\ee 
The latter relation indicates transparently the discrete nature of the appropriate group spaces \footnote{The origin of Eq. (\ref{dl}) was revealed in the monograph \cite {Al.S.}, in \S T20.

It is the particular case of the more general relation
$$  \pi_i (K)=  \pi_i (L_1)+\dots + \pi_i (L_r) 
    $$ 
for a group $K$ which is the  product of the groups $L_1 \dots L_r$ at a fixed $i$ (it is correctly for  the Lie groups of the series $SU$, $U$ and $SO$, with which modern theoretical physics deals). This works good for the Abelian gomotopy groups $\pi_0$ and $\pi_1$, we study in the present article.  }.

\medskip
To construct the degeneration space (vacuum manifold) $R_{\rm YM}\equiv G/H$ corresponding to the  discrete representation (\ref{fact2}) for the gauge group  $H$, we utilize 
 Eq. \cite {rem1,Ryder}
\be 
\label{sxema1} 
G=H \oplus G/H \equiv H \oplus R.\ee 44
We must substitute  in (\ref{sxema1})  the "discrete" representation (\ref{fact2}) for the appropriate gauge symmetries groups:
 \be
 \label{trick}
  G_0= {\bf Z}\otimes U_0 + G_0/(U_0\otimes {\bf Z}).
 \ee  
 From  latter Eq. we learn that the degeneration space in the Minkowskian Higgs model with vacuum BPS monopoles quantized by Dirac \cite {Dir} is 
 \be
 \label{RYM}
 R_{\rm YM}=  G_0/(U_0\otimes {\bf Z}). \ee

\medskip

Our reasoning in the previous section showed that {\it domain walls are impossible} for such vacuum manifold. However various rotary effects still allowed for the vacuum manifold (\ref{RYM}), as we shall convince ourselves soon. 

 At our study of the Minkowskian YMH model with vacuum BPS monopole solutions, quantized by Dirac and involving nontrivial topological dynamics, it is very seductive also to develop the version of this model {\it involving} domain walls and maintaining herewith  nontrivial topological dynamics.  

Let us now discuss whether such YMH model with domain walls can be constructed. We here propose the following idea. As in the  ’t Hooft-Polyakov theory \cite{Polyakov,Ryder, Cheng},  we assume the "standard" way violating the initial $SU(2) $ gauge symmetry, namely $SU(2) \to U(1)$. But we already saw  \cite{Pervush1} that the gauge group $ U(1)$ permits the "discrete" factorization 
\be \label{f11} U(1)\cong U_0\otimes {\bf Z} \equiv \tilde U, \ee
with $ \pi_0 (U_0)=\pi_1(U_0)=0$. This factorization can be justified, as we have discussed above, by utilizing the "loop framework" \cite{Postn4} embodied in Eq. (\ref{cll}). Indeed the manifold $\tilde U$ can be treated as the disjoint union of the topological
sectors $U_1^n$ ($n\in {\bf Z}$), (\ref{disj}). Then also the schematic proof of the isomorphism (\ref{f11}) was given.

It is very tempting, looking at it, to assume the following further way violating the initial $SU(2)$ gauge symmetry:
\be \label{3way}  SU(2)\to U(1)\to \tilde U.
\ee
Here an important question arises: how legal to treat $\tilde U$ as the degeneration space (vacuum manifold) for the gauge symmetry break down (\ref{3way})?  The argument in favor of this fact is that $\tilde U$ proves to be invariant with respect to the gauge group $U(1)$ thanks to the homotopical class $\tau$ which just once uniformly running around
the circle  $S^1$and herewith counterclockwise.

The principal result which we need it now is that domain walls {\it exist} inside $\tilde U$, $\pi_0 \tilde U={\bf Z}$! 

\medskip Note that the same way violating the initial $U(1)$  symmetry and with the same disjoint union $\tilde U$, {\it which we assume to be also gauge}, inherent in the topological model for helium ${\rm He}^4$. As we argue in the present study, it is necessary to justify superfluid phenomena (corresponding to the topologically trivial sector  $U_1^0$) and rotary effects (Khalatnikov vortices \cite{Halatnikov}) in the topologically nontrivial sectors  $U_1^n$. This implies, in a natural way, the break down the initial (gauge) $U(1)$ symmetry of the model with appearance   of domain walls inside the appropriate  vacuum manifold $\tilde U$.  The case is for the experimenters and phenomenologists to check this hypothesis!

\medskip Above we have discussed in detail the manifest invariance of the YMH model with vacuum BPS monopole modes and nontrivial topological dynamics with respect to going over from any topology $k$ to its opposite, $-k$. We have shown, in this context, that this involves the modification 
$$U_0 \otimes {\bf Z} \rightarrow U_0 \otimes {\bf Z} \otimes {\mathbb Z}_2   $$ 
in constructing the "modified" vacuum manifold $R'_{YM}$. But this modification  does not change anything in the topological considerations for $R'_{YM}$ in comparison with $R_{YM}$ since $ {\bf Z} \otimes {\mathbb Z}_2 \cong  {\bf Z}$. In particular, once again, as it is easy to see, there are no domain walls inside $R'_{YM}$!

\bigskip
The nontrivial isomorphism (\ref {iso}) \cite{Al.S.} satisfying (as we  have discussed above) for the vacuum manifold $R_{YM}$, (\ref{RYM}),  implies the presence of thread topological defects inside this manifold. \par 

\bigskip On the level of phenomenology this implies \cite{Al.S.}  violating  the thermodynamic equilibrium along definite lines in the given  vacuum manifold. \par
In the simple case when such lines, threads,  are straight, the thermodynamic equilibrium is violated in an infinitely narrow tube around each of these threads. Thus regions of thread topological defects have the topology of a cylinder (topologically equivalent to the circle $S^1$). But it is well known that $\pi_1 S^1={\bf Z}$.\par
In our researches about the Minkowskian Higgs model with vacuum BPS monopoles quantized by Dirac \cite{Dir} we assume that the region of thread topological defects inside the vacuum manifold: $R_{YM}$ versus $\tilde U$, is the infinitely narrow tube of the infinite length around the  axis $z$ in the chosen rest reference frame. The effective "diameter" of this cylinder of the infinite length is 
just $O[\epsilon(\infty)]$, that is, indeed, an infinitely small value.  \par
In the next section we shall  ground the above assumption investigating the properties \cite{Al.S.}  of rectilinear  threads $A_\theta $ inside the vacuum manifold. \par  
\medskip
Furthermore, it   may be shown that
\be \label {pointd}
\pi_2(R_{YM})= {\bf Z}= \pi_1 H.\ee
We have already discussed the isomorphism $\pi_1 (H)= {\bf Z}$ with the aid of the exact sequence (\ref{fibr}).

In this point of our discussion we would like to give the alternative argument proving this isomorphism.

This is the arguments of the  holonomies group, stated in Refs. \cite{LP2, Al.S.}. 
Actually, these come to the existence of one-dimensional loops in the $H$ group space, forming  holonomies elements
  \be
 \label{cycl}
 b_\gamma =P\exp (-\oint _{\Sigma} {\bf T} \cdot A _\mu dx^\mu)
\ee 
over one-dimensional cycles $\Sigma$.\par  
The  holonomies elements $ b_\gamma$, belonging to the $U(1)\subset SU(2)$ embedding, make the complete holonomies group (we shall denote also as $H$) isomorphic to $U(1)$ \cite{LP2}. \par  
Eq. (\ref{cycl}) makes the isomorphism $\pi_1 (H)= {\bf Z}$ highly transparent (as that coming to $\pi_1 S^1={\bf Z}$).\par  
The     relation
$$  \pi_2(R_{\rm YM})= \pi_1 H,    $$ 
is the particular case of  more   general Eq.  (\ref{hedge}). \par  
The good  proof of the isomorphism (\ref{hedge}) was demonstrated in the monograph \cite{Al.S.}, in \S T20, and we recommend our readers this monograph to acquaint   with it. \par 

\medskip Let us verify explicitly, with the aid of the appropriate exact sequence, that this "general" isomorphism has place to be for the vacuum manifold $R_{\rm YM}$ in its shape (\ref{RYM}).

This exact sequence (which is the continuation of the  exact sequence (\ref{lsec1})) has the look 
\bea \label{lsec2} \ldots \longmapsto \pi_2(U_0\otimes {\bf Z}) \longmapsto \pi_2(SU(2)) \longmapsto \pi_2(SU(2)/(U_0\otimes {\bf Z})) \\ \nonumber \longmapsto \pi_1(U_0\otimes {\bf Z}) \longmapsto \pi_1(SU(2)) \longmapsto \pi_1(SU(2)/(U_0\otimes {\bf Z})) \longmapsto \pi_0(U_0\otimes {\bf Z}) \ldots
\eea
But 

$\bullet$ $\pi_2 (SU(2))=0$ since $SU(2)\cong S^3$ while $\pi_2 S^3=0$ \footnote{See Lectures 26 and 27 in \cite{Postn4} for the sphere $S^n$ and its homotopical groups.}. 

$\bullet$ Also $\pi_1 (SU(2))=0$, for the same reasons  (in other words, $SU(2)$ is one-connected).

$\bullet$  $\pi_2(U_0\otimes {\bf Z})=0$ since $U_0\otimes {\bf Z}\cong U(1)\cong S^1$ and $\pi_2 S^1=0$.

$\bullet$ $\pi_2(R_{\rm YM})=\pi_2(SU(2)/(U_0\otimes {\bf Z})) = \pi_1 (U_0\otimes {\bf Z})=\bf{Z}$. The latter relation can be read off from  the exact sequence (\ref{lsec2}). Really,  $\pi_2 (SU(2))=\pi_1 (SU(2))=0$ and also $\pi_1(U_0\otimes {\bf Z})$ as it was argued above. This gives the desired result!

\medskip On the other hand, from the same exact sequence (\ref{lsec2}), $\pi_2(U_0\otimes {\bf Z})=0$ and this {\it prohibits point topological defects} in the 
$\tilde U$,  (\ref{disj}). It is an evident result now.  In other words, it {\it doesn't suit us} as the way break-down the initial $SU(2)$ gauge symmetry. And thus we stop, in our study, on the option of the vacuum manifold $R_{\rm YM}=SU(2)/(U_0\otimes {\bf Z})$, i.e. that without domain walls but possessing thread and point (hedgehog) topological defects! 

\medskip 
The nontrivial  isomorphism (\ref{pointd}) testifies in favour of point hedgehog topological defects inside the vacuum manifold $ R_{\rm YM}$ in the Minkowskian Higgs model with vacuum BPS monopoles quantized by Dirac \cite{Dir}.\par  
These  point hedgehog topological defects and  vacuum BPS monopole solutions are   indivisible.  \par  
As in the case of   (vacuum)  't Hooft-Polyakov monopole solutions \cite{H-mon, Polyakov}, associated with the "continuous"  vacuum $\sim S^2$ geometry  (\ref {contin}) and the FP "heuristic"  quantization scheme   \cite{ FP1}, point hedgehog topological defects inside the discrete vacuum manifold $ R_{\rm YM}$, (\ref{RYM}), come to violating the thermodynamic  equilibrium in an infinitesimal neighbourhood of the origin of coordinates. \par  
In particular,  this    violating  implies the singularity of the appropriate vacuum  "magnetic"   fields $\bf B$, serving as the order parameters in the mentioned Higgs models, at the origin of coordinates.   \par
In spite of definite distinctions between the 't Hooft-Polyakov and BPS Higgs models (these distinctions were pointed out, for instance, in the recent paper \cite{rem1}; the principal of them is the absence of superfluid properties in the 't Hooft-Polyakov model \cite{H-mon, Polyakov}) the   vacuum  "magnetic"   field $\bf B$ diverges as $r^{-2}$  at the origin of coordinates in both models \footnote{for the 't Hooft-Polyakov model \cite{H-mon, Polyakov} this was demonstrated in Ref. \cite{Ryder}, while for the BPS Higgs model, in the original papers \cite{BPS}.}:
\be  \label{radial pole1} B_k\sim \frac{r_k}{gr^3}. \ee
\medskip
Moreover,  there is a likeness between Eqs.  (\ref{pointd})  and (\ref{point1})  as  the particular cases of the general relation  (\ref {hedge}) \cite{Al.S.}, inducing point topological defects in (gauge) theories. \par
On the other hand,   there are no thread  topological defects inside the continuous vacuum manifold  $R$,   (\ref {contin}), since $\pi_1 S^2=0$ and moreover when the "continuous"  geometry is assumed for the residual $U(1)$ gauge symmetry in the (Minkowskian) Higgs model, also $\pi_0 ~U(1)=0$.\par
\bigskip  

Now we would like to prove some very important result, namely that 

\be \label{subset11} S^1\subset R_{\rm YM}\cong SU(2)/(U_0\otimes {\bf Z}).
\ee  There is a natural projection 
\be \label{prn} \pi: \quad  SU(2) \to  R_{\rm YM}\cong SU(2)/(U_0\otimes {\bf Z}).
\ee  Subtract now a maximal torus $T\cong U(1)\subset SU(2)$ in $SU(2)$:  $$T \cong U(1)\subset SU(2). $$ The example of such torus is that consisting on the diagonal matrices of the type 
\be \label{diagonal}T= \begin{pmatrix} e^{i\theta} & 0\\ 0 & e^{-i\theta} \end{pmatrix} \ \Big|\ \theta\in[0,2\pi) \cong S^1. \ee

\medskip Now recall that $$ \pi_0 (U_0)= \pi_1 (U_0)=0,$$ i.e. that $U_0$ is connected and simple connected. Then its direct product with $\bf Z$ is responsible for different topological sectors in the YMH vacuum!

In this context, it is possible to choose the maximal torus $T\subset SU(2)$ in such a way that

\be \label{sect}  T\cap  (U_0\otimes {\bf Z}) =\{ e\},
\ee i.e. that the "continuous", $U_0$, as well as "discrete", $\bf Z$, parts of $U_0\otimes {\bf Z}$ give only the trivial contribution in this intersection with this concrete subtorus.

Note that such possibility (\ref{sect}) is the direct consequence of two factors. Firstly, because   $ \pi_0 (U_0)= \pi_1 (U_0)=0,$ i.e. because $U_0$ is connected and simple connected. Secondly,  because discrete elements from $\bf Z$  can be chosen don't belong to $T$  (or  at least in such a wise that their intersections with  $T$ would be trivial).

\medskip Let us consider now the restriction of the projection $\pi$ on the torus $T$,
\be \label{pit}  \pi|_T:~ T\to R_{\rm YM}. 
\ee
Since the kernel of $ \pi|_T$  is

\be \label{kerpit} 
{\rm ker} (\pi|_T)= T\cap  (U_0\otimes {\bf Z})=\{ e\},
\ee  then $ \pi|_T$  is an embedding (more precisely, it is an {\it injective} local homeomorphism onto the image). Thus, the image

\be \label{image}  S^1\equiv \pi(T)\subset  R_{\rm YM}
\ee is an embedded circle: 

\be \label{embc}\pi(T) \cong T \cong  S^1.
\ee  Thus the concrete embedding $S^1\to  R_{\rm YM}$ is obtained. This allows to speak correctly about the loops  $\Gamma_n \subset S^1\subset  R_{\rm YM}$, with the homotopical classes given via  (\ref{cll}):
$$ [\Gamma_n]=[u]^n =[e_{p_0}] \circ \tau^n,$$  with $\tau$ being the homotopical class of loops which just once uniformly running around
the circle  $S^1$ and herewith counterclockwise.  Indeed,  $\tau$ is the generator of the fundamental group $\pi_1 S^1={\bf Z}$  \cite{Postn4}. 

\bigskip  Why do we need this reasoning?  The point is that in the paper \cite{rem1}, the theory of magnetic charge was stated in fact for the more simple case when the vacuum manifold in the YMH model with magnetic monopoles is isomorphic to the sphere $
S^2$.

It's quite natural to ask a question now, how much the picture changes if we replace the vacuum manifold $R\cong S^2$ (for instance in the ’t Hooft-Polyakov theory) with the vacuum manifold $ R_{\rm YM} \cong SU(2)/(U_0\otimes {\bf Z})$ in the discussed YMH model?

Recall now the general formula for the magnetic charge $\bf m$ \cite{rem1,Al.S.}:

\be \label{m.zarjad}  {\bf m}_\Gamma  = \frac 1{4\pi} \int _\Gamma B~ds
\ee over the two-dimensional surface $\Gamma \subset R^3$.

In \cite{rem1}, the most calculations about the magneti1c charge touched on the two-dimensional homotopic group $\pi_2 S^2$, with $S^2$ being indeed $S^2_\infty$.   This corresponded to the vacuum manifold 
$$ R\equiv SU(2)/U(1) \cong S^2. $$

In this framework,  the magnetic charge $\bf m$  proves to be directly proportional to the degree of map  $\pi_2 S^2$. For details, we refer our readers to \cite{rem1,Al.S.}.  Important points in the justification of this magnetic charge theory were:

the continuous geometry $R\cong S^2$; this allows to chose   smooth surfaces $\Gamma$ and expose them by continuous deformations;

the flux remains invariant at continuous deformations of the  smooth surface $\Gamma$  (until  this flow intersects with the core of the monopole!).

\medskip In the present study, the picture is changing!   Although in (\ref{lsec2})  we  proved that $\pi_2  (R_{\rm YM})={\bf Z}$,  this is not an unique important result  (although also important, of course)!
 Herewith the topological classification shifts now to the  groups $\pi_0$  and  $\pi_1$. 

\medskip  The principal problem with the vacuum manifold $R_{\rm YM}$  is that there are not a smooth $S^2$ structure (as in the case of the vacuum manifold $R$ in the 't Hooft-Polyakov model): the "continuous"  part of $R_{\rm YM}$ is topologically trivial while the topologically nontrivial nature of this manifold is associated with the group $\bf Z$ of integers.

Nevertheless, Eq. (\ref{m.zarjad}), as the definition of the magnetic charge ${\bf m}_\Gamma$, remains absolutely correct if 

$\bullet$ $\Gamma$  is a closed surface in $R^3$  covering the monopole. 

$\bullet$  Linearizing the Bianchi identity

$$ D_{[iF_{jk}]} =0\longleftrightarrow D_i B^i=0; \quad B_i =\frac 1 2 \epsilon_{ijk} F^{jk},$$  i.e. going over to the effective {\it Abelian}  field onto the large distances, one can speak about the "physical" magnetic field  $\vec B_{\rm phys}$  as  about a vector in $R^3$  with the usual divergence $\nabla \cdot  \vec B_{\rm phys}$.

In this framework, $\nabla \cdot  \vec B_{\rm phys}=0$  outside  the core of the YM BPS monopole  \footnote{We recommend our readers the original paper by E.B. Bogomol’nyi in \cite{BPS} where the manifest look of the "magnetic" field $\bf B$ is given.}.   In other words, we can write down 

$$ \nabla \cdot  \vec B_{\rm phys}=0  \quad {\rm on}~~ R^3/{\rm kernel}.  $$

\medskip  Thus one can conclude that the local differential part of our proof (namely, YM equation and linearized Bianchi identity  $\nabla \cdot  \vec B_{\rm phys}=0$ outside the singularity) regardless of whether  $S^2\cong R$ is continuous or $R_{\rm YM}\cong SU(2)/(U_0\otimes {\bf Z})$ is discrete.  The only thing which  is changing is the topological justification of the magnetic charge ${\bf m}_\Gamma$ (more exactly, its quantization way).

\medskip  How this justification changes in the "discrete" vacuum geometry framework?  Now the quantization  of the magnetic charge ${\bf m}_\Gamma$ is not associated with the degree of map for $S_\infty^2 \to R$ but with other things, namely with the facts that:

$\bullet$  the wave functional must be  single-valued  (or be transformed with a fixed  phase) at the transition between the sectors $n\to n+1$;

$\bullet$ this results the quantization condition similar to the Dirac quantization.
2
In this framework, the flux $\int _\Gamma B dS$ is interpreted as the "difference" between the sectors.  In the discrete geometry, it is quite naturally to 
interpret the magnetic charge as an invariant  measuring  the transition between neighboring  vacuum sectors. 

\medskip  To sum it up, the definition (\ref{m.zarjad}) of the magnetic charge as the flux  must be preserved. Also we preserve the local arguments, in particular, the linearized Bianchi identity  $\nabla \cdot  {\bf B}=0$ outside the core of monopole and independence  from choice the surface $\Gamma$ in the physical space. 

 \medskip As a {\it modification}, we can propose  to consider the $\Gamma \subset R^3$ surface and associate the magnetic charge ${\bf m}_\Gamma$ with the discrete structure $\pi_0(R_{\rm YM})={\bf Z}$ and "large"/ "small" gauge transformations.  Note herewith that this magnetic charge ${\bf m}_\Gamma$ is 
closely related to the topological number $n$ via the behavior of YMH fields at the spatial infinity.

\medskip Thus our "modus operandi" is following:

$\bullet$  The first step. A loop $\Gamma_n$ is selected with the homotopic class $[u]^n$.

$\bullet$ The second step. The action of the appropriate "large"/ "small" gauge transformation onto the wave functional  is considered.

 $\bullet$  The claim of the single-value nature of this wave functional results the quantization condition connecting $n$ with   ${\bf m}_\Gamma$: ${\bf m}_\Gamma \propto n$.

Then 

$\bullet$ $\Gamma_n$ can be interpreted as a  marker of the appropriate topological sector;

$\bullet$  ${\bf m}_\Gamma$  is the physical magnetic charge measuring via the flux through  $\Gamma  \subset R^3$;
  
$\bullet$  the connection between these two things is carried out by means of quantum conditions 	imposed onto the "large"/ "small" gauge transformations.

\medskip Let us now write it down carefully in which a way the loops $\Gamma_n$ (and appropriate "large"/ "small" gauge transformations!)  enter the quantization condition for the magnetic charge. 

Consider once again the vacuum manifold $R_{\rm YM}\cong SU(2)/(U_0\otimes {\bf Z})$ with 

$$ \pi_0  (U_0)= \pi_1  (U_0)=0 $$  and where the discrete factor $\bf Z$  encodes the set of topologically different vacuum sectors.  It is quite naturally to parametrize these sectors by integers $n \in {\bf Z}$.

As we have proved above,  $S^1\subset R_{\rm YM}$. Such $S^1$ can be treated as an effective circle arising as the image of the one-dimensional submanifold in the space of the "large"/ "small" gauge transformations  (related to the residual $U(1)$ gauge group!).  Since due to (\ref{clla}),

$$[u]^n\cong [\tau] ^n,$$  the loop $\Gamma_n$  implements the $n$
-multiple going around the circle  $S^1$. Thus the integer-valued index $n$  of the vacuum sector can be identified naturally with the number going the loop $\Gamma_n$ around the circle  $S^1$.

\medskip  Consider now the "large" gauge transformations corresponding to the loops $\Gamma_n$. Let us denote temporally as $g_n$ the element of the (residual) gauge group associated with $\Gamma_n$.  Its action onto the gauge field $A$  and the wave functional is set in the standard way as 

\be \label{dejstvie}  A\mapsto  A^{g_n}; \quad \Psi[A]\mapsto \Psi^{(n)}[A]=\Psi[ A^{g_n}]. 
\ee 
	In the framework of the Dirac quantization of the magnetic charge in the YMH theory, it is required for the  the wave functional to be unambiguous  (or, more generally, quasiperiodical with a fixed phase)  under the action of "large"/ "small" gauge transformations. This condition can be written down as
	\be \label{phas}\Psi[ A^{g_n}]=e^{i\alpha n}\Psi [A]
	\ee  where $\alpha$  is a constant determined by the physical parameters of the model (in particular, by the magnetic charge). 
	
	On the other side, the action of the gauge transformation  $g_n$  can be associated with  changing the topological functional (for instance, one can talk about the Chern-Simons functional or its effective analogue in the studied model), which, in turn, is expressed via the flux of the (effective) "magnetic"  field through the closed surface  $\Gamma \subset R^3$:
	\be \label{potok} {\bf m}_\Gamma = \frac 1{4\pi} \int _\Gamma  B dS.
	\ee
	
	Then the phase entering Eq. (\ref{phas}) can be written down in the shape 
	\be \label{phas1}\alpha n  = q{\bf m}_{\Gamma} n,
	\ee  where $q$  is the effective coefficient dependent on the normalization of the gauge field and the charges  present in the theory. Herewith  the condition  for the  the wave functional to be unambiguous at $n\to n+1$  requires that
	
	$$ e^{i\alpha } =e^{iq{\bf m}_{\Gamma}}=1,$$ i.e.
	\be \label{quant} q{\bf m}_{\Gamma}= 2\pi k;  \quad k \in {\bf Z}.
\ee  From (\ref{quant}) it follows the quantization of the magnetic charge ${\bf m}_{\Gamma}$:
\be \label{mquant}
{\bf m}_{\Gamma}=\frac {2\pi} q  k ;  \quad k \in {\bf Z}.
\ee
Thus, the integer index $n$	realizable via the loops $\Gamma_n$  in $S^1\subset R_{\rm YM}$  and the integer index $k$ defining  the quantization of the magnetic flux ${\bf m}_{\Gamma}$ turn out to be connected via the phase of the wave functional under the action of "large"/ "small" gauge transformations.  Unlike the case of the "continuous" vacuum geometry $SU(2)/U(1)\cong S^2$, where  the quantization of the magnetic flux ${\bf m}_{\Gamma}$  is derived mostly from $\pi_2 S^2$, in the "discrete" vacuum geometry  $R_{\rm YM}\cong SU(2)/(U_0\otimes {\bf Z})$,  the quantization of the magnetic flux ${\bf m}_{\Gamma}$  is  relied on the discrete structure $\bf Z$ of the vacuum sectors and the homotopical classes of loops $\Gamma_n \subset S^1$.
	

\bigskip 
In the next section  it will be argued that the first-order phase transition occurs indeed in the Minkowskian Higgs model with vacuum BPS monopoles   quantized by Dirac \cite{Dir}. \par
\medskip It will be very useful, in this context, to discuss an useful analogy which us give the liquid helium II model. \par
Just in the latter model, possessing manifest superfluid properties \cite {Landau}, also the first-order phase transition happens. 
It is associated with (spontaneous) violating the superfluidity   \cite {Landau} in a  (resting) liquid helium II specimen along definite (rectilinear) lines, threads.  
This implies  \cite {Halatnikov} the presence of thread (rectilinear) vortices in that specimen. \par
The similar situation with arising (rectilinear) vortices takes place when  this specimen  turns together with the cylindrical vessel in which it is contained.   \par 
The phenomenology of the latter case was stated enough good in the monograph \cite{Landau52}, in \S 29. We don't intend, in the present study, to retail all the said in the monograph \cite{Landau52} about this subject; however we, for all that, would  like to elucidate some theses of the helium turning model. This will be very helpful for us   in the next section. \par
A  good argument in favour of the first-order phase transition occurring in the liquid helium II turning model is the nontrivial contribution $\Delta E$ \cite{Landau52} in the total helium (He$^3$) Hamiltonian \footnote{ This has the typical look \cite {Nels} 
\begin{eqnarray}
\widehat{\cal H} &=& \int d^3r~\hat a^+ ({\bf r},t)\left({-\hbar^2\over 2m}\;\nabla^2\right)
\hat a({\bf r},t)\nonumber\\
&\quad +&{1\over 2}
\int d^3r\int d^3r'\hat a^+
({\bf r},t)~\hat a^+({\bf r}',t) V(|{\bf r}-{\bf r}|)~
\hat a({\bf r}',t)~\hat a({\bf r},t). \nonumber 
\end{eqnarray}
  Boson creation and annihilation operators: $ a^+({\bf r},t) $ and $ a^+({\bf r},t)$, respectively, may be expressed in terms of the phase $\Phi (t,{\bf r})$  of the Bose condensate wave function (\ref {Xi1}) \cite{Landau52}.
\par
Note that such expressing  boson creation and annihilation operators in terms of the one phase function $\Phi (t,{\bf r})$  is in a good agreement with the quantum-mechanical calculations about the liquid helium II proposed by Bogolubov and co-authors \cite{N.N.}, in particular, with
the Bogolubov transformations \cite{N.N.,Levich, Smir}
$$ \hat b^+ _{{\bf p}} =u_{{\bf p}} \hat \xi ^+_{{\bf p}} + v_{{\bf p}}\hat \xi  _{{\bf -p}},   $$
with 
$$ u_{{\bf p}}^2- v_{{\bf p}}^2=1 $$
and the creation (respectively, annihilation) operators 
$$
\hat b^+ _{{\bf p}} = \frac{\hat  a _0 \hat a^+ _{{\bf p}}}{\sqrt{n_0}},
 \quad \hat b _{{\bf p}} = \frac{\hat a^+ _0 \hat a _{{\bf p}}}{\sqrt{n_0}},  $$
expressed through the "initial" creation (annihilation) operators entering actually the Bogolubov model Hamilton operator, may be represented as \cite{N.N.,Levich}
$$ \hat H= -\sum \limits_{a=1} ^N \frac{\hbar^2}{2m} \Delta_a +\frac{1}{2}  V (\vert {\bf r}_a -{\bf r}_b \vert),   $$
(with $V (\vert {\bf r}_a -{\bf r}_b \vert)$
being the  interaction energy between particles $a$ and $b$ and  $N$ being the complete number of  particles in the considered system),
brought in the diagonal form
$$  \hat H=\hat H_0+ \sum \limits_{ {\bf p} \neq 0} \epsilon ({\bf p})
\hat \xi ^+ _{{\bf p}}\hat \xi  _{{\bf p}}   $$
via the {above Bogolubov transformations}.

Herewith  the creation (respectively, annihilation) operators $ \hat a^+ _0$ and $\hat a_0$ corresponds to the zero momenta ${\bf p}=0$ and the number $n_0$ of helium atoms possessing these zero momenta.} from the global solid \it potential \rm
rotations  of the helium specimen turning together with the cylindrical vessel where it is contained.

This contribution is given with Eq.
\be \label{vklad1} \Delta E \sim L \pi \rho_s \frac{\hbar ^2}{m^2} \ln \frac{R}{a}:
\ee
with $L$ being the length of the vessel, $\rho_s$ being the density of the superfluid component in the helium II specimen, $m$ being the mass of a helium atom; at last $R$ and $a$ are, respectively, the radius of the vessel and an arbitrary distance of atomic scales.\par
The shortcoming of this Eq. is its  logarithmic divergence at $R/a\to\infty$.

\medskip
The important peculiarity  of the liquid helium II turning model \cite{Landau52} is the appearance of rectilinear threads (accompanied by appropriate vortices),  parallel  to the rotation axis (altitude) of the cylindrical vessel (herewith chosen to coincide with the axis $z$ of the given rest reference frame). 

\medskip
The   liquid helium II turning model \cite{Landau52} possesses initially the manifest   $U(1)~\otimes ~O(2)$  symmetry.

In  this model $ O(2)$ is the group of {\it global} (rigid) potential rotations around the axis $z$  of the liquid helium II specimen and the  cylindrical vessel where it is contained.  




\medskip
As it is well known, superfluidity phenomena \cite{Landau} in a liquid helium II  specimen are associated with violating the initial $U(1)$ gauge symmetry of the Bogolubov  Hamiltonian $\widehat{\cal H}$.
This second-order phase transition occurs in the fixed Curie point $T_c\to 0$.

To explain in this case  rotary effects (coming to appearance of thread vortices) in the liquid helium II turning model \cite{Landau52} (as well as in the liquid helium II at rest theory \cite{Halatnikov}), it is worth to assume that violating the initial $U(1)$ gauge symmetry in the liquid helium II theory in such a wise  that the $U(1)\simeq S^1$ group space turns into the discrete quite separated set \cite{Engel} \footnote{The notion of quite separated sets \cite{Engel} means in this case  that always there exists such a   function $f: \tilde U\to I$ ($I\equiv [0,1]$) for two homotopical classes $A$ and $B$ inside $\tilde U$ that $f(x)=0$ as $x\in A$, while $f(x)=1$ as $x\in B$. Here $x$ is a (one-dimensional) way.  
One speaks in this case that $f$ \it separates \rm the sets $A$ and $B$.}. 

More exactly,  it may be assumed the spontaneous breakdown 
\be \label{subgr} U(1) \simeq U_0 \otimes {\bf Z}\longrightarrow  \tilde U(1) \simeq {\bf Z}; 
\quad \pi_0 U_0 =\pi_1 U_0=0.\ee
Geometrically,  the discrete (quite separated) set $\tilde U(1)$, us discussed above, may be seen as the circle $S^1$ cut off by its topological domains (vice verse, to get the continuous $U(1)$ group space, these topological domains would be again pasted together); see our explanation above. A remarkable fact in this context is that $\tilde U(1)$ has the properties of the degeneration space, i.e. it is invariant with respect to the $U(1)$ (gauge) group inherent in the ${\rm He}^4$ model.  

\medskip
On the other hand, in the liquid helium II turning model \cite{Landau52}, the appropriate group $O(2)$ of rigid rotations remains exact.

This implies the (spontaneous) breakdown
\be \label{O2} O(2) \otimes U(1) \longrightarrow O(2) \otimes \tilde U(1)\simeq O(2) \otimes U_0 \otimes  {\bf Z}\equiv H  \ee
of the initial symmetry group $ O(2) \otimes U(1) $ inherent in the liquid helium II turning model \cite{Landau52}.

\medskip
Upon some mathematical manipulations, resembling somehow ones \cite{Ryder} led to Eq. (\ref{RYM}) in the Minkowskian Higgs model with vacuum BPS monopoles   quantized by Dirac, the  degeneration space in the liquid helium II turning model \cite{Landau52} may be founded. It proves to be
  \be
 \label{R1}
 [{\bf Z}\otimes O(2)] \otimes U_0\equiv R_{\rm turn}. \ee 

Recall herewith  that the group $O(2)$ of $2\times 2$ 
orthogonal matrices with determinants $\pm 1$ always may be represented as   
$$ SO(2) \otimes {\Bbb Z}_2,$$
with $ SO(2)\simeq U(1)$ being the group of orthogonal matrices with determinants $+1$, while ${\Bbb Z}_2\subset \bf Z$. 

It can be argued that $\pi_1 R_{\rm turn}=\pi_1 S^1={\bf Z}$.  These arguments come to the nature of {\it global} $O(2)$ rotations. This group $O(2)$ consists, as we have noted above, of the subgroup ${\mathbb Z}_2=\{1,-1\}$ and global $SO(2)$ rotations. 

Begin now with ${\mathbb Z}_2$. Let us consider how it influences the liquid ${\rm He}^4$ specimen turning around the axis $z$ of the cylindrical vessel.  Note, with this purpose, that the topological numbers  $n\in {\bf Z}$ proves to be \cite{Halatnikov} \be \label{zirk11} n=\frac m{2\pi \hbar} \oint _\Sigma {\bf v}^{(n)}; \quad m/\hbar ~~{\rm is ~fixed}.\ee

Here linear velocities vectors $ {\bf v}^{(n)} $ in each topological class $n$ can be chosen (for instance, by means of a Galilean transform) in such a way that $\vert{\bf v}^{(n)}\vert=\vert {\bf v}_0\vert $ independently on a concrete topological
charge $n$.  These $n$ can be treated naturally as winding numbers (Chern-Simons functionals) in the ${\rm He}^4$ topological model. 

In this context, the action of  ${\mathbb Z}_2$ onto the helium specimen comes to the transformations $n\to \pm n$, in other words, {\it a vortex maps onto itself or into its ant-vortex}! We still have an infinite number of topological sectors marked with integers, but just now there is a symmetry that “turns over” them. It is remarkable at the same time that the trivial topological sector $n=0$ involving the superfluid helium component maps onto itself; in other words, global $O(2)$ rotations {\it do not change superfluid
properties of helium}!

\medskip 

As a consequence of the said for the group $H$, (\ref{O2}), we have \footnote{To calculate the number of connection components, i.e. the group $\pi_0$ of the tensor product $H$, utilize the mentioned already simple rule for the $\pi_n$ of the tensor product of (topological) groups. For instance, for $\pi_0$ and $n=3$, $$ \pi_0 (G_1\otimes G_2\otimes G_3)=\pi_0 (G_1) \otimes\pi_0 (G_2) \otimes \pi_0 (G_3)$$
(and so for each $\pi_n$ and any number of groups).
Also we take into account that  $\pi_0 U_0=0$ and $\pi_0 {\bf Z}={\bf Z}$.} \be \label{H0} \pi_0 H ={\bf Z} \otimes {\mathbb Z}_2 \cong {\bf Z}. \ee

\medskip It remains still to study the  influence the global $SO(2)\simeq U(1)$ rotations. These rigid $SO(2)$ rotations on the some {\it constant} angle $\alpha$ do not change the Bogolubov Hamiltonian $\hat {\cal H}$.  This is so since the operator $\hat {\cal H}$ contains the mutually Hermitian-conjugated creating/annihilation operators $\hat a^\dagger ({\bf r},t)$,  $\hat a ({\bf r},t)$, respectively.  So the global $SO(2)$ rotation, involving the exponent $\exp(i\alpha)$ do not alter the products of the $\hat a^\dagger \hat a$ type.   Note in this context that the rigid $SO(2)\cong U(1)$ symmetry saves the number of (quasi)particles in the specimen. This, as it can be proven, does not change the thermodynamical properties of ${\rm He^4}$. 
 
Thus $R_{\rm turn}$ is topologically equivalent to 
\be \label{rturn}S^1\cong U(1)\cong U_0\otimes {\bf Z}\cong \tilde U.\ee   This is correctly modulo global $O(2)$ transformations, as we have just shown!

Therefore we can speak that the degeneration space $R_{\rm turn}$ inherits all the topological properties of $\tilde U$. Namely, 

\be \label{rturn-top} \pi_0 R_{\rm turn} =\pi_1 R_{\rm turn} ={\bf Z}.\ee

Now, from ({\ref{H0}), (\ref{rturn-top}) we read off the relation 
\be \label{verteksu} \pi_0 H= \pi_1 R_{\rm turn}={\bf Z},\ee which guarantees the existence of vortices, i.e. thread topological defects, inside $R_{\rm turn}$. Thus the general isomorphism (\ref {iso}) \cite{Al.S.} is satisfied in the case of the liquid helium II turning model \cite{Landau52}.



In this concrete case it implies the presence of (rectilinear) thread topological defects (vortices) in the quested model.
These rectilinear vortices contribute with  the item $\Delta E$ \cite{Landau52}, (\ref {vklad1}), additional to the Bogolubov helium (diagonalized) Hamiltonian $\widehat{\cal H} $ \cite{Nels}.

Just this increasing $\Delta E$ the helium energy (in comparison with  that given by $\widehat{\cal H} $ and referring to the helium specimen at rest) is the sign of the first order phase transition occurring. Recall herewith that the group $O(2)$ of global rotations does not change the termodynamical properties of liquid ${\rm He}^4$!

\medskip
Also, there are  walls between different topological domains inside the discrete liquid helium II turning degeneration space $R_{\rm turn}$, (\ref{R1}), because of Eq (\ref{rturn-top}).

On the other hand, there are no point topological defects inside $R_{\rm turn}$ since $\pi_2 S^1=0$. 

\bigskip
 The case when a liquid helium II is at rest is somewhat simpler than the case \cite{Landau52} of the liquid helium II turning together with  the cylindrical vessel where it is contained.

In the former case, to justify the spontaneous appearance of rectilinear vortices \footnote{As it was explained in \cite{Halatnikov} (see also \cite{ rem3}), the appearance of rectilinear vortices in a liquid helium II specimen is set by Eq.
$$ n= \frac{m}{2\pi \hbar} \oint \limits _\Gamma {\bf v}^{(n)}d {\bf l}; \quad n \in {\bf Z}. $$
This Eq. implicates the helium mass $m$ and the tangential velocity ${\bf v}^{(n)}$ of a rectilinear vortex.

In this case, trivial topologies $n=0$ correspond to disappearing the cyclic integral on the right-hand side of latter Eq. It is just the case of superfluid potential motions in the given liquid helium II specimen with 
$ {\rm rot}~ {\bf v}^{(0)}=0$.

In Ref. \cite{ rem3}, the explanation of vortices \cite{Halatnikov} in a (rested) liquid helium II specimen as a particular case of the {\it Josephson effect} \cite{Pervush3} (coming to circular persistent motions of material points) was given. }, it may be presumed that the initial  $U(1)$ {\it gauge} symmetry group of the Bogolubov helium Hamiltonian $\widehat{\cal H} $ \cite{Nels} is violated in the 
\be\label{tilde} U(1)\to \tilde U \ee
wise.

As in the liquid helium II turning case \cite{Landau52}, 
$\pi_0 ~\tilde U  ={\bf Z}$.

On the other hand, it may be demonstrated (applying the manipulations similar to those necessary in the liquid helium II turning case \cite{Landau52} to ascertain the look (\ref{R1}) for the appropriate degeneration space $R_{\rm turn}$) that the degeneration space $\tilde R$ in the liquid helium II at rest theory coincides with the residual gauge symmetry group 
$$R_r\cong \tilde U\cong U_0\otimes {\bf Z}.$$ 
Thus, formally,
$$  \pi_1 \tilde R = \pi_0 ~\tilde U  ={\bf Z}, $$
and this just implies the existence of thread topological defects (rectilinear vortices) in the liquid helium II at rest theory. Simultaneously, the relation $\pi_0 ~\tilde U  ={\bf Z}$ provides domain walls inside $\tilde R$.

\medskip The alone fact  \cite{ rem3, Halatnikov} referring vortices in a liquid helium II at rest to nontrivial topologies $n\neq 0$, while superfluid potential motions are referred to trivial topologies $n= 0$, is very remarkable.



Thus the connection between the  superfluid and rotary effects in the liquid helium II model and the topological degeneration of appropriate data  may be observed.

Finally, it may be argued, as in the  liquid helium II turning model \cite{Landau52}, that the first-order phase  transition occurs also in the liquid helium II at rest model \cite{Halatnikov}.

\bigskip It is interesting to compare our results about the liquid helium II  and those gotten in the monograph \cite{Volovik}. 

More exactly, in the  monograph \cite{Volovik}, the group of {\it rigid} space rotations $SO(3)_{\bf L}$ around the axes $x$, $y$, $z$ (involving the orbital momentum ${\bf L}$) for a ${\rm He}^4$ specimen (instead of $O(2)$ in our case) \footnote{The mentioned rigid $SO(3)_{\bf L}$ is the natural group of rigid space rotations in the 3-dimensional Euclidian space which does not change the Bogolubov Hamiltonian $\hat {\cal H}$, the number of (quasi)particles and the thermodynamical properties of liquid ${\rm He}^4$. Rigid rotations of a liquid ${\rm He}^4$ specimen together with the cylindrical vessel where it is contained \cite{Landau52}, involving the group $O(2)$, is an additional construction. } was considered. As the author \cite{Volovik} asserts, the appearance of vortices course violating there the initial $U(1)$ symmetry involves also violating $SO(3)_{\bf L}$ since the direction of the (one) vortex line {\it appears as the axis of spontaneous anisotropy}. This anisotropy line can be chosen as the axis $z$ of the (fixed) reference frame.

\medskip In our case, with the (rigid) $O(2)$ symmetry group, it is a somewhat different situation since we already have  two-dimensional (rigid) rotations about the fixed axis, which, in a natural way,  can be chosen as the axis $z$.
Thus it is now an already ready anisotropy (and simultaneously {\it symmetry}) line, and there are no need for violating $O(2)$ in this case, and the above 'topological' calculations (\ref{O2}) remain legitimate.

\medskip The author  \cite{Volovik} considers a complex scalar $\Psi=\vert\Psi \vert \exp(i\Phi)$ as the order parameter for the superfluid $^4$He. Note that the similar look for the order parameter (coinciding with the helium
Bose condensate wave function) was proposed in the monograph \cite {Landau52}:
\be
\label{Xi11}
 \Xi (t,{\bf r})= \sqrt {N(t,{\bf r})}~ e^{i\Phi(t,{\bf r})},
\ee
with $ N(t,{\bf r})$ being the number of particles in the ground energy state $\epsilon=0$ (we shall continue the analysis of the helium
Bose condensate wave function in Section 4).\par

The different from zero order parameter $\Psi$ implies the complete breakdown of the $U(1)_{N}$ symmetry. 

For a vortex with the winding number $n_1$, it was set \cite {Volovik} $\Phi(t,{\bf r})\equiv n_1\phi$.

It is natural to guess  \cite {Volovik} that the symmetry-breaking scheme in the presence of  $SO(3)_{\bf L}$ is 
\be \label{scheme}
G^{'} = U(1)_{N} \otimes SO(3)_{\bf L}\rightarrow {\bf H}^{'}=U(1)_Q
\ee
Here the remaining symmetry group $U(1)_Q$ is the symmetry with the order parameter given in (\ref{Xi1}). It is the rotation by the angle $\theta$ that transforms $\phi \to \phi +\theta$, accompanied by the global phase rotation $\Phi\to\Phi +\alpha$, with $\alpha=-n_{1}\theta$. The generator of such $U(1)_Q$ transformations is
\be \label{qgen}
Q=L_{z}-n_1 N.
\ee

\medskip From the said in \cite{Volovik} we see a principal difference of the G. Volovik approach to liquid helium ${\rm He}^4$ with that developed in the present study. Our analysis is based upon the assumption  about the initial {\it gauge} $U(1)$ symmetry of the Bogolubov helium Hamiltonian $\hat {\cal H}$, violated then in the (\ref{subgr}) wise.  As for the {\it global} $SO(3)_{\bf L}$ symmetry, which is present naturally in the 3-dimensional flat Euclidian space, we can "forget" about this symmetry which does not change the model. 

On the other hand, global vortices corresponding to the $U(1)_Q$ group ala \cite{Volovik} {\it also should present}. This can lead to forming composite vortex structures. Also a global vortex may induce a local gauge response. A local vortex may inherit angular momentum from the global sector via the 
$ Q$
 generator.

In the sphere of the stability the discussed theory, the following conclusions can be drawn. Local vortices have finite energy due to gauge screening. Their coexistence leads to nontrivial energy landscapes, possibly stabilizing bound vortex states or vortex lattices.  All this should be the subject of  an additional study!

\medskip Note that the same phenomenological assumptions can be made also for global $O(2)$ vortices in the liquid helium-4 placed in the cylindrical vessel turning around its axis $z$. As it was discussed above, such global $O(2)$ vortices are connected with rotations on  constant angles $\alpha$. Such rotations form a subgroup in $O(2)$.

\bigskip
Returning again to the Minkowskian Higgs model with vacuum BPS monopoles   quantized by Dirac \cite{Dir}, note the very transparent analogy between the rotary item $\Delta E$, (\ref{vklad1}), in the  liquid helium II turning model \cite{Landau52} and the free rotator action 
functional $W_N$, (\ref {rot}), in the former case.

The important distinction between the both theories is, however, in the logarithmic divergence of $\Delta E$, (\ref{vklad1}), as $R/a\to\infty$, while the free rotator action 
functional $W_N$, (\ref {rot}), is suppressed actually by the value of $\epsilon(\infty)\to 0$.

\medskip
The just performed brief analysis of the liquid helium II turning model \cite{Landau52} (as well as of the liquid helium II at rest model) suggests, by analogy, the idea to link the  various rotary properties \cite{ rem3, David2,LP2,LP1, Pervush2} of the Minkowskian Higgs model to vacuum BPS monopoles   quantized by Dirac (described correctly by the free rotator action 
functional  (\ref {rot})) with the discrete vacuum geometry (\ref{RYM}), including thread topological defects inside the appropriate vacuum manifold $R_{\rm YM}$ (in the similar way, has been outlined above, in which such topological defects induce rotary effects in the liquid helium II theory).

This will be the topic of the next section, where also some arguments in favour of the first-order phase  transition occurring in the Minkowskian Higgs model with vacuum BPS monopole solutions   quantized by Dirac and similar to that taking place in the liquid helium II theory (these arguments are connected with  the "electric" and "magnetic" contributions to the total vacuum energy).
\section{How topologically nontrivial threads arise inside the discrete non-Abelian vacuum manifold.}
The outlines of this topic were us projected in Introduction.

It should be begun from the fact that the Minkowskian Higgs model \cite{Al.S., BPS, Gold} with vacuum BPS monopoles is the specific model where the duality in specifying the sign of the vacuum "magnetic" field $\bf B$ takes place. This duality is just set by the Bogonol'nyi equation (\ref{Bog}) \cite{ Gold}.

This duality implies automatically the duality in specifying the signs of magnetic charges $\bf m$, given by the general formula \cite{Al.S.}
\be
\label{mul} 
{\bf m} =\frac {1}{4\pi} \int\limits _\Gamma d{\bf S}~ {\bf B} = \frac {1}{8\pi}\int d^3 x ~\partial _i \{\epsilon ^{ijk} <F_{jk}^b,\Phi_b >a^{-1}\}.
\ee 
Note that latter Eq. is irrelatively correct, actually, to the concrete choice of the vacuum geometry: either "continuous" or "discrete" one, in the quested Minkowskian Higgs model.

Really,  the definition (\ref {mul}) of magnetic charges $\bf m$ may be recast formally to the look
´\be \label{magnutnuj zahjad}  {\bf m} = \frac {1}{4\pi} \int\limits _\Gamma d{\bf S}~ {\bf B} \sim \frac {1}{4\pi} \int\limits _V {\rm div}~ {\bf B}~ dV,    \ee

and this integral does not depend on  the choice of  the space-like surface in the (Minkowski) space. On the other hand, as it follows from this equation, if 

$$ <B>\equiv \int_V d^3 x B =0,  $$ this does not implies ${\bf m}=0$. It is an important conclusion we widely utilize in the present study.\par
 However, the actual divergence of the "magnetic" tension  ${\bf B}$ at the origin of coordinates in all the  Minkowskian Higgs models (for instance, \cite{BPS,H-mon,Polyakov}) implies  that the latter integral is different  from zero as  the origin of coordinates lies inside the chosen space-like surface $\Gamma$.\par 
Furthermore, the above duality \cite{Gold} in specifying the sign of the vacuum "magnetic" field $\bf B$ affects also the generator $h(\Phi)$ \cite{Al.S.} of the residual $U(1)$ gauge group in the studied Minkowskian Higgs model. \par
It is so since $h(\Phi)$ may be read easily from Eq. (\ref{mul}).\par
\medskip 
Now let us return to the simpler case of "continuous" ($\sim S^2$) vacuum geometries in Minkowskian Higgs models. \par 
In this case the appropriate degeneration space: say, $R$, is one-connected:  $$\pi_0 (H)= \pi_1(R)=0,$$ 
and one  can always choice, in the unique way,  a continuous branch of the function $ h(\Phi)$ on $R$.\par
Another situation is when the degeneration space $R$ is multi-connected ("discrete"), i.e. when  
$$\pi_0 (H)= \pi_1(R)\neq 0,$$ 
and therefore there exist thread topological defects in the gauge theory in question (in particular, in the Minkowskian Higgs model with vacuum BPS monopoles quantized by Dirac \cite{Dir}). \par
On the one hand (as we have seen this above), the nontrivial isomorphism (\ref{iso}) \cite{Al.S.} implies the existence of thread topological defects inside the discrete vacuum manifold $R$.\par 
On the other hand, it turns out (see \S $\Phi$13 in \cite{Al.S.}) that the duality in the choose of the sign of $ h(\phi)$ (and therefore also the duality \cite{Gold} in the choose  of the direction of  $\bf B$)  may become unavoidable. \par 
This occurs, in particular, when one can invert the  function $ h(\Phi)$ via the gauge transformations 
\be \label{dvaznaka} \tilde h(\Phi) =\alpha h(\Phi) \alpha^{-1}=- h(\Phi),\ee
with
\be 
\label{alef}
 \alpha =\exp (2\pi M)
\ee  
and  the element $M$ belonging to the Lee algebra of the  group $G_1$ of  global transformations  compensating changes in the vacuum (YM-Higgs) configuration $(\Phi^a,A_\mu^a)$ at rotations around the  axis $z$ of the chosen reference frame. \par 
The general look of an element $g_\theta\in G_1$ may be set by Eq. (\ref{gteta}) \cite{Al.S.}. \par 
Herewith our above discussion about thread topological defects inside degeneration spaces in the liquid helium II models (turning \cite{Landau52} and at rest \cite{Halatnikov}) prompts us the idea to associate the rectilinear threads (as infinite thin tubes of  infinite lengths)  with the above axial symmetry. \par 
In the monograph \cite{Al.S.} (in \S $\Phi$12)  the strict proof was given that the unitary group $G_1$ is a possible  symmetry of alike threads (by analogy with the group $O(2)$ of orthogonal $2\times 2$ matrices with determinants $\pm 1$,  one utilizes in the liquid helium II turning model \cite{Landau52}). \par 
Omitting details, note  that the proof of this statement comes to the proof  that in each topological class  rectilinear threads exist giving a finite contribution to the appropriate energy integrals \rm (due to first-order phase transitions   occurred\rm)  and satisfying the appropriate equations of motion \rm  \cite{Al.S.}. \par
The topological type of a thread defect is determined by the group $G_1$: more exactly by its generator $M$\rm. \par 
To  prove  this statement, we betake to the argument of the holonomies group \cite{LP2, Al.S.}. \par 
Let us rewrite an element (\ref {cycl}) of the holonomies group $H\simeq U(1)$ in the shape 
\be  \label{holonomija1} \alpha= (P\exp(-\oint A_\mu dx^\mu ))^{-1}. \ee 
Due to the standard Pontryagin degree of a map theory,  elements $\alpha$  of the holonomies group $H$ depend on integers $n$: $\alpha\equiv \alpha(n)$ ($n\in \bf Z$). \par 
Passing then to the cylindrical coordinates, we come to Eq. (\ref{holonomija})  \cite{ Al.S.},  implicating rectilinear (topologically nontrivial) threads $A_\theta$, may be represented as (\ref {Ateta}).\par 
Alternatively, $A_\theta$ may be represented also as  \cite{ Al.S.}
\be
  \label{Aro}
 A  _\theta (\rho) = M+ \beta (\rho),
\ee 
where the function $\beta (\rho)$ approaches zero as $\rho \to \infty$.  \par 
This also ensures \cite{Al.S.} a finite energies densities in  Minkowskian Higgs models involving  (topologically nontrivial) threads. \par 
Thus at the spatial infinity elements $ \alpha (n)\in H$ acquire the look 
\be 
\label {holonomija2}
\alpha(n)= \exp (2\pi M). 
\ee 
We see that the dependence of  gauge fields $A$ on the angle $ \theta$ disappears at the spatial infinity. \par  
It is equivalent to  damping, at the spatial infinity,  various rotary effects, associated in Minkowskian 
non-Abelian theories with thread topological defects. \par
In particular, it is correctly for the  Minkowskian Higgs model with vacuum BPS monopoles quantized by Dirac \cite{Dir} and involving the "discrete" vacuum geometry (\ref{RYM}).\par
On the other hand, the function $\beta (\rho)$ should be different   from zero effectively only in the $\epsilon$-neighbourhood of the  origin of coordinates to ensure in this infinitesimal region the above pointed collective rotary phenomena in the appropriate Minkowskian YM vacuum via the nontrivial dependence of $ A  _\theta (\rho, \theta, z)$ on $\theta$. \par 
\medskip
Indeed, above described (vacuum) YM modes $ A  _\theta (\rho, \theta, z)$ \cite{Al.S.} would be different from vacuum YM BPS monopoles (for instance, in the Minkowskian Higgs model with vacuum BPS monopoles quantized by Dirac). \par
It is associated with the obvious divergence as $r^{-2}$ \cite{BPS} (c.f. (\ref{radial pole1}) \cite{Ryder}) of the vacuum "magnetic" field $\bf B$ at the origin of coordinates in Minkowskian non-Abelian models involving vacuum  BPS monopole solutions. \par
In this case there is impossible to link YM fields $ A _ \theta (\rho, \theta, z)$, representing topologically nontrivial (rectilinear) threads in the Minkowskian Higgs model to vacuum BPS monopoles quantized by Dirac, with vacuum YM BPS monopole solutions (with same topological numbers), at least in a smooth wise (that is connected with the $\sim \partial_\mu A_\nu$ items always entering the "magnetic" field $\bf B$ in YM theories).\par
On the other hand, the value $<B^2>$, the vacuum expectation value (VEV) of the "magnetic" field $\bf B$ squared, can serve \cite{LP1}  as the order parameter in Minkowskian Higgs models with YM fields. \par
Herewith its nonzero value, $<B^2>\neq 0$, is the sign violating the initial $SU(2)$ gauge symmetry group down to its $U(1)$ subgroup, and it is the second-order phase transition \cite{rem1,Linde}.\par

The natural question arise here. Why $<B^2>$ but not the VEV of the squared Higgs field (in the shape of a Higgs BPS monopole mode) {\it serves in our case as the order parameter}? The explanation is the following \cite{LP1}. It turns out that the Higgs (effective) mass $m/\sqrt{\lambda}$  (where $m$ and $\lambda$ are the Higgs mass and self-interacting constant, respectively) goes to infinity in the limit $V\to \infty$ at assuming that $<B^2>$ is finite in this limit. That's why.  It (this effective mass) may be given as  \cite{ David2,LP2,LP1} 
\be 
\label{masa}  \frac{1}{\epsilon}=\frac{gm}{\sqrt{\lambda}}\sim \frac{g^2<B^2>V}{4\pi}. \ee 
Thus $\epsilon$  is, in turn, inversely proportional to the spatial volume  $V=\int d^3 x$ occupied by
the appropriate (YM-Higgs) field configuration. This volume is {\it fixed}.

The said allows to assert that $\epsilon$  (and $\sqrt \lambda/m$ in turn) disappears as $1/V$ at the spatial infinity in
the infinite spatial volume limit  $V\to\infty$. On the other hand, in the zone of asymptotic
freedom of quarks at the origin of coordinates, when the YM (gluonic) constant $g$ goes
to zero, $\epsilon$ {\it can take any finite values} (due to the  $(0\times \infty)^{-1}$ uncertainty in the case of the
fixed infinite volume).

 Thus \cite{LP1} the scalar (Higgs) field acquires an infinitely large mass and disappears from the spectrum
of physical excitations. And the role of the order parameter of the physical BPS monopole vacuum is "transmitted" to $<B^2>$ in this infinite volume limit.

\bigskip If the Minkowskian Higgs model with YM fields implies the continuous $\sim S^2$ vacuum geometry, the possible (as in the BPS or 't Hooft-Polyakov models) divergence of $\bf B$ at the origin of coordinates testifies \cite{rem1} in favor of point hedgehog topological defects located just at the origin of coordinates. \par 
The origin of such topological defects is in the nontrivial isomorphism (\ref{hedge}).\par 
This singularity of (vacuum) "magnetic" fields can be removed by regularising appropriate energy integrals. \par
Also there are no contributions from point hedgehog topological defects into these  energies integrals unlike the case of first-order phase transitions when such contributions appear (as, for instance, $\Delta E$ \cite{Landau52}, (\ref {vklad1}), in the case of a liquid helium II specimen turning together with  the cylindrical vessel where it is contained). \par
\medskip Unlike the former case, gaps in the directions of $\bf B$ and ${\bf B}_1$ in the Minkowskian YM-Higgs model with vacuum BPS monopoles quantized by Dirac and 
involving the "discrete" vacuum geometry (\ref{RYM}) according to our assumption are evidences in favour of the first-order phase transition occurring in this model. \par 
This is in a good agreement with the general theory of first-order phase transitions (discussed, for instance, in Ref. \cite{Linde}, in \S 3.1; see also \cite{rem1}).  \par 
The discontinuous behaviour of plots for order parameters is just the sign of first-order phase transitions occurring in physical theories. \par 
Indeed, to prove that the  first-order phase transition takes place in the, we discuss now, Minkowskian YM-Higgs model with vacuum BPS monopoles quantized by Dirac, a large job is necessary. In particular, it is necessary to analyze the first derivatives of the thermodynamics potentials in the both mentioned thermodynamics phases: the "rotary" and "superfluid" inside the YMH  vacuum quantized by Dirac.  In the present study we do not set ourselves as the goal the such analysis, but nevertheless we would like make the one important notice already here.

A some insight in the analysis of first-order phase transitions prompts us that it should be

\be \label{B1v}
<B_1^2> =0,
\ee
i.e. one can  presume that the "rotary" component of the YMH  vacuum quantized by Dirac can be interpreted as a "false" vacuum. We emphasize that this is a preliminary assumption, need a further research: checking that at $<B_1^2> =0$ the free energy of the YMH  vacuum quantized by Dirac reaches a minimum and that this minimum is not the absolute one, but this is beyond the scope of this article.  

Due to Eq. (\ref{B12}),  the "false" vacuum condition (\ref{B1v}) comes (roughly) to the condition
$$\partial_ \rho A_\theta (\rho,\theta, z) =0.$$

\medskip
The above discussed gaps do not influence however the "right" of gauge matrices $ \alpha (n)$, (\ref {holonomija}) (taking account of the  "discrete" vacuum geometry (\ref{RYM}), involving [topologically nontrivial] threads), to represent the holonomies group $H$ coinciding with the residual gauge symmetry group $U(1)$ in the  Minkowskian Higgs model with vacuum BPS monopoles quantized by Dirac \cite{Dir}.\par 

This can be explained by the natural isomorphism between exponential multipliers $ \alpha (n)$, referring to (topologically nontrivial) threads arising because of the "discrete" vacuum geometry (\ref{RYM}), and Gribov topological multipliers $v^{(n)} ({\bf x})$, referring to  superfluid potential motions in the  above model. \par\medskip 




\bigskip Note   that the  dependence of the  theory considered on nontrivial topologies $n$ also disappears at the spatial infinity, i.e. in the infrared region of transferred  momenta. \par 
It is the next in  turn evidence in favour of the infrared topological confinement (in the spirit \cite{fund,Azimov}) occurring in the  Minkowskian Higgs model with vacuum BPS monopoles quantized by Dirac. \par 

\bigskip
The above discussed examples of liquid helium II at rest  \cite{Halatnikov} and turning together with  the cylindrical vessel where it is contained \cite{Landau52} give us  pattern of physical systems in which first-order phase transitions occur. \par 
The "center of gravity" of the enumerated models is in the  appearance of (rectilinear) threads generated appropriate vortices, i. e. thread topological defects. \par
On the other hand, it was demonstrated above (repeating the arguments \cite{Landau52}), in the case of a liquid helium II specimen turning together with  the cylindrical vessel where it is contained, the "rotary" contribution $\Delta E$, (\ref {vklad1}), supplements the typical (diagonalized) Bogolubov Hamiltonian $\widehat{\cal H}$ \cite{Nels}.\par
It is the sign of the first-order phase transition taking place in the liquid helium II turning model \cite{Landau52}.\par 
Herewith according to the modern terminology \cite{ rem1}, this "rotary" contribution $\Delta E$, (\ref {vklad1}), is the {\it latent heat}, while the thermodynamic phase  in whose framework collective solid potential rotations \cite{ Landau52,Halatnikov} in a liquid helium II specimen turning together with  the cylindrical vessel where it is contained coexist with superfluid motions setting by the Hamiltonian $\widehat{\cal H}$ is characterized by the {\it supercooling} phenomenon. \par 
In this case superfluid motions inside a liquid helium II specimen represent the {\it stable state} corresponding to the {\it second-order} phase transition occurred therein and coming to violating the $U(1)$ gauge symmetry of the helium Hamiltonian $\widehat{\cal H}$ \cite{Nels} \footnote{The transparent evidence in favour of the { second-order} phase transition occurring in a helium: actually, it is the transition between the liquid helium I high-temperature and liquid helium II modifications (the latter one possesses manifest superfluid properties described in the Landau-Bogolubov theory \cite{Landau,N.N.}) is the discontinuity \cite{Levich1} in the plot of the helium heat capacity in the Curie point $T_c\to 0$.\par 
For instance \cite{Nels}, in the case of superfluid helium films, the experimental data result
$$ C_p(T)\equiv T(\frac{\partial S}{\partial T})\approx c_1+c_2\exp[c_3/\vert T-T_c\vert^{1/2}]    $$
for the helium heat capacity $ C_p(T)$,
with $S$ being the appropriate entropy; $c_i$ ($i=1,2,3$) are constants. } (in the (\ref {O2}) wise in the liquid helium II turning case \cite{Landau52}).\par
In turn, collective solid potential rotations in this specimen are referred to the {\it metastable state} (in the terminology \cite{Linde}).\par 
\medskip
There are concrete computations, given in Ref. \cite{Nels} with the example of a rested liquid helium II specimen, demonstrating discontinuities in the plot of the vacuum expectation value $<\vert \Xi\vert ^2 (t,{\bf r})>$ for the helium Bose condensate wave function $ \Xi$, (\ref {Xi1}),
 near locations of vortices cores. \par
Herewith the value $< \vert \Xi \vert ^2 (t,{\bf r})>$ can serve quite naturally as the order parameter in the Landau-Bogolubov helium theory \cite{Landau,N.N.}. \par 
So $< \vert \Xi\vert ^2 (t,{\bf r})>=0$ for the helium I modification \cite{Levich1}, corresponding to the initial $U(1)$ gauge symmetry of the helium Hamiltonian $\widehat{\cal H}$ \cite{Nels}, while $<\vert\Xi\vert^2 (t,{\bf r})>\neq 0$ upon this symmetry is violated in the (\ref{subgr}) wise, that corresponds to the helium II modification. \par 
\medskip
Then, utilizing Eq. (\ref {Xi1}) \cite{ rem1, Landau52}, the correlation (Green) function
$$  G(r)\equiv<\Xi(t,{\bf r}),\Xi^*(t,{0})> $$ 
(it is different from zero for the liquid helium II modification) may be recast to the look \cite{Nels}
\begin{eqnarray}
G(r)&=&\langle\Xi(t,{\bf r})\Xi^*(t,0)\rangle\nonumber \\
&=& \Xi_0^2\exp\left[-{1\over 2}\langle[\Phi(t,{\bf r})-
\Phi(t,0)]^2\rangle\right].
\label{eq:seventeen}
\end{eqnarray}
To a first approximation, we can neglect fluctuations in the amplitude of $\Xi$, setting it to be a $\Xi_0$.
  On the other hand,  the above discussed expression \cite{Halatnikov}
for topological indices $n\in {\bf Z}$ via circular velocities ${\bf v}^{(n)}$ is mathematically equivalent to  \cite{Nels}
\begin{equation}
\frac 1 {2\pi }\oint_{\Gamma}\nabla \Phi ~ d{\bf l}=
\int_\Omega d^2r\;n_v({\bf r}),
\label{omega}
\end{equation}
with the vortex "charge density" given as
\begin{equation}
n_v({\bf r})=\sum_{\alpha=1}^N\;
n_\alpha\delta({\bf r}-{\bf r}_\alpha)
\label{eq:thirtyone}
\end{equation}
for a collection $N$ of vortices
located at positions $\{ {\bf r}_\alpha \}$ with integer charges $\{n_\alpha\}$.  Herewith the contour $\Gamma$ enclosing many "elementary" vortices such that
 \be
\oint\;\Phi\cdot d {\bf l}=
2\pi s_\alpha\;.
\label{twentynine}
\ee  with $s_\alpha=\pm 1$.  Then we can replace the sum of "elementary" vortices by an one vortex with the fixed topology $N\in \bf Z$. 

Without loss of generality, it may be set $z=0$ in all the calculations \cite{Nels} following.
In (\ref{omega}), Eq. (\ref{alternativ}) \cite{rem1,Landau52}, with appropriate replacing $ {\bf v}_0\to{\bf v}^{(n)}$, was utilized.

Applying  then the Stokes formula to   (\ref{omega}), one find upon integrating:
\begin{eqnarray}
\epsilon_{ij}\partial_i\partial_j\Phi (t,{\bf r})&=&
\partial_x\partial_y\Phi-\partial_y\partial_x\Phi\nonumber \\
&\approx& n_v({\bf r}).
\label{eq:thirtytwo}
\end{eqnarray}
To cast this Eq. in a
more familiar form, one can introduce \cite{Nels} the value $\tilde \Phi(t,{\bf r})$ dual to $\Phi(t,{\bf r})$:
\begin{equation}
\partial_i\Phi(t,{\bf r})=\epsilon_{ij}\partial_j
\tilde\Phi(t,{\bf r}).
\label{eq:thirtythree}
\end{equation}
Then
\begin{equation}
\nabla^2\tilde \Phi(t,{\bf r})=n_v({\bf r}).
\label{eq:thirtyfour}
\end{equation}
In particular,
\be 
\label{triv}
\nabla^2\tilde \Phi(t,{\bf r})=0
\ee 
in the case when vortices are absent, i.e. in the case of purely superfluid motions  inside a liquid helium II specimen.

Eq. (\ref{eq:thirtyfour}) is the particular case of the Poisson equation. Thus it permits its
\begin{equation}
\tilde\Phi(t,{\bf r})=\sum_{\alpha}\;n_\alpha G({\bf r},{\bf r}_\alpha)
\label{eq:thirtyfive}
\end{equation}
solution, where the Green function $ G({\bf r},{\bf r}_\alpha)$  (defined as in the left-hand side of Eq. (\ref {eq:seventeen}))  satisfies
\begin{equation}
\nabla^2 G({\bf r},{\bf r}_\alpha)=\delta({\bf r}-{\bf r}_\alpha).
\label{eq:thirtysix}
\end{equation}
For $\vert{\bf r}-{\bf r}_\alpha \vert $ large enough and both points far from any boundaries, the Green function $ G({\bf r},{\bf r}_\alpha)$  has the look (see \S11.8 in \cite{V.S.Vladimirov})
\begin{equation}
G({\bf r},{\bf r}_j)\approx{1\over 2\pi}\ln\left(
{|{\bf r}-{\bf r}_j|\over
\xi_0}\right)+C
\label{eq:thirtyseven}
\end{equation}
on the plane $z=0$,
where $C$ is a constant which contributes to the vortex core energy.

The origin of the parameter $\xi_0$ entering Eq. (\ref{eq:thirtyseven}) is following \cite{Nels}.

It is the characteristic length related to the coefficients of the first two terms of Eq. \cite{Landau52}
\begin{equation}
{F\over T}=
\int\; d^2r\left[{1\over 2}\;A|\nabla~\Xi|^2+
{1\over 2}\;a|\Xi|^2+b|\Xi|^4+
\cdots\right]
\label{eq:eleven}
\end{equation}
for the helium free energy $F$.

In this case $\xi_0$ may be defined as \cite{Nels}
\be \label{xio}
\xi_0=\sqrt{A/\vert a\vert },
\ee
with $a$ given as $a=a'(T-T_c)$.

As it follows from (\ref{eq:thirtyseven}), at setting $C=0$, the Green function $ G({\bf r},{\bf r}_j)$ diverges logarithmically when  $\vert{\bf r}-{\bf r}_j \vert \to\infty$, while it can approach zero when 
$$ \vert{\bf r}-{\bf r}_j \vert \to\xi_0.$$
Additionally, $ G({\bf r},{\bf r}_j)$ diverges logarithmically also at $\vert{\bf r}-{\bf r}_j \vert \to 0$. The latter property of $ G({\bf r},{\bf r}_j)$, as we shall se soon, is very important. 

\medskip The just performed analysis of  $ G({\bf r},{\bf r}_j)$  allows to draw the series important conclusions about the phase $\tilde\Phi(t,{\bf r})$ of the helium wave function  $\Xi$ (taking over the complete collection $N$ of topologies: including the trivial one $s_\alpha=0$, corresponding to superfluid motions \cite{Landau,Halatnikov}, in a liquid helium II specimen at rest).

So, due to (\ref {eq:seventeen}), 
\be \label{srednee}
<\vert (\Xi) \vert ^{2}(t,{\bf r})> \equiv <\Xi (t,{\bf r}),\Xi^*(t,{\bf r})>\approx G(0)=1/N \sum\limits _\alpha < \vert \Xi^{2} \vert _\alpha (t,{\bf r}) >,\ee
serving the order parameter in the helium  theory, diverges logarithmically at each index $\alpha (n)$ ($n\in{\bf Z}$) in  the same (\ref{eq:thirtyseven}) \cite{Nels} sense that the Green function $ G({\bf r}_\alpha,{\bf r}_\alpha)
\approx G(0)$, i.e. in the points $\{{\bf r}_\alpha \}$ where quantum  vortices are located in a liquid helium II specimen at rest \footnote{In (\ref {srednee}) the Green function $G(0)$ is different from that given via Eq. (\ref {eq:seventeen}) and involving the constant amplitude $\Xi_0$.

Now the amplitude of $\Xi (t,{\bf r})$ wouldn't be constant, and this implies that $G(0)$ has the nontrivial look $ G({\bf r}_\alpha,{\bf r}_\alpha)$ \cite{Nels}, (\ref {eq:thirtyseven}).}.

This is associated immediately with (rectilinear) quantum vortices arising spontaneously \cite{Halatnikov} in a liquid helium II specimen at rest.

Herewith it is not important that also the phase $\tilde \Phi(t,{\bf r})$ (or $\Phi(t,{\bf r})$) of the helium wave function  $\Xi$ possesses the same behaviour that the Green function $ G({\bf r},{\bf r}_j)$.

The said is correctly for nontrivial vortices topologies $n\neq 0$ due to Eq. (\ref {eq:thirtyfive}). For $n=0$, i.e. in the "superfluid case", $\tilde \Phi(t,{\bf r})$ becomes an uncertain value when $\vert{\bf r}-{\bf r}_j \vert \to\infty$ (or when $\vert{\bf r}-{\bf r}_j \vert \to 0$).



\medskip
Summarizing, now one can assert that in the presence of (rectilinear) quantum vortices \cite{Halatnikov} in a liquid helium II specimen, the order parameter $<\vert \Xi\vert ^2(t,{\bf r})>$ in the helium model suffers discontinuities of the logarithmic nature in the points $\{{\bf r}_\alpha \}$ in which these (rectilinear) quantum vortices are located (the long distances logarithmic singularities of $ G(0)$ are, probably, not associated with quantum vortices).

The discovered discontinuity (of the logarithmic nature) in the plot of the vacuum expectation value $<\vert \Xi\vert ^2 (t,{\bf r})>$ for the helium Bose condensate wave function $ \Xi$, serving the order parameter in the helium theory, just testifies in favour of the first-order phase transition occurring in a liquid helium II rested specimen with arising therein quantum vortices.  

There are, indeed, lot of distinctions between the Minkowskian Higgs model with vacuum BPS monopoles quantized by Dirac \cite{Dir} and the liquid helium II at rest theory \cite{Halatnikov} (the same concerns also the liquid helium II turning model \cite{Landau52}, involving rigid $O(2)$ rotations).

But, in spite of these distinctions, the discontinuities in the appropriate order parameters:  $<\vert \Xi\vert ^2(t,{\bf r})>$ in the helium theory and $<B^2>$ in the Minkowskian Higgs model with vacuum BPS monopoles quantized by Dirac, can be considered as the common trait of the both models associated with rotary effects (including the "discrete" vacuum geometry and thread topological defects).

Moreover, drawing a parallel between the Minkowskian Higgs model with vacuum BPS monopoles quantized by Dirac and the liquid helium II at rest theory \cite{Halatnikov}, one can refer  collective solid rotations inside the appropriate BPS monopole vacuum to the metastable thermodynamic phase, characterised by the latent heat  equal to the "electric" energy $\sim E^2$. This "electric" energy is described correctly by the free rotator action functional (\ref {rot}).

Simultaneously, superfluid potential motions inside this vacuum, involving the "magnetic" energy $\sim B^2$, may be referred to the stable thermodynamic phase.

It is again the supecooling situation.

\medskip 
As it was discussed in  Section 2, on the face of it, there is a contradiction between the rotary effects in the Minkowskian Higgs model with vacuum BPS monopoles quantized by Dirac (these effects, including collective vacuum rotations, are determined by the action functional (\ref{rot})) and the manifest potential nature of the Minkowskian BPS monopole vacuum. 

More exactly, this potential nature of the Minkowskian BPS monopole vacuum (suffered the Dirac fundamental quantization \cite{Dir}) comes, in particular, to the colinearity (\ref{colin}) \cite{Pervush1} of vectors $\bf B$ and $\bf E$ (vacuum "magnetic" and "electric" fields, respectively).

This is associated with the explicit look (\ref {el.m}), (\ref {se}) \cite{David2,LP2,LP1, Pervush2} for vacuum "electric" monopoles $\bf E$, can be expressed through vacuum Higgs BPS monopole modes $\Phi_{(0)}({\bf x})$.

Also the vacuum "magnetic" field can be expressed through vacuum Higgs BPS monopole modes $\Phi_{(0)}({\bf x})$, now through the Bogomol'nyi equation (\ref{Bog}). 

\medskip
There is however a way to solve the above contradiction. 

It turns out that, side by side with (topologically nontrivial) threads $A_\theta$, one can construct also specific  vacuum Higgs $z$-invariant solutions \cite{Al.S.}, having the look (\ref{Higgs-teta}) in the cylindrical coordinates, such that the vacuum Higgs field $\Phi_a^{(n)}$ specified by (\ref{Higgs-teta}) merges with the appropriate Higgs vacuum BPS monopole solution in the Minkowskian Higgs model with vacuum BPS monopoles quantized by Dirac \cite{Dir}, herewith belonging to the same topology $n$. \par
This may be achieved by the appropriate choice of the function $\phi(\rho)$ in (\ref{Higgs-teta}). \par
Additionally, we should also claim that the covariant derivative $D\Phi$ of any vacuum Higgs field $ \Phi_a^{(n)}$ specified via (\ref{Higgs-teta}) merges with the covariant derivative of such a vacuum Higgs BPS monopole solution. \par
In other words, vacuum Higgs solutions (\ref{Higgs-teta}), corresponding to the above described axial symmetry $G_1$, would merge with appropriate Higgs vacuum BPS monopole solutions actually in a smooth wise.
\par 
By this the goal is pursued to link $D\Phi_a^{(0)}$  to vacuum "electric" monopoles (\ref {el.m}) in the Minkowskian Higgs model with vacuum BPS monopoles quantized by Dirac. \par 
Indeed, this confluence of vacuum Higgs solutions would occur already at the notably lesser scale of distances: at the hadronic distances scales, $\sim$1 fm., to ensure the correct infrared behaviour \cite{Azimov} of quark Green functions in the Minkowskian (constraint-shell) QCD. \par
\medskip
Actually, the following picture takes place. \par 
As it was noted in Ref. \cite{David3}, Higgs BPS monopole solutions disappear at the origin of coordinates (in its infinitesimal neighbourhood of the effective radius $\epsilon$). \par 
In this situation the continuous and smooth prolongation of Higgs vacuum BPS monopole solutions with appropriate Higgs modes $ \Phi_a^{(n)}$ \cite{Al.S.}, (\ref{Higgs-teta}), seems to be quite natural. It is controlled by the proper choice of $\phi(\rho)$.\par 
\medskip
It is just the way solving the above discussed problem how to co-ordinate the "colinearity condition" (\ref{colin}) \cite{Pervush1} (reflecting the superfluid nature of the Minkowskian BPS monopole vacuum suffered the Dirac fundamental quantization \cite{Dir})  with collective solid rotations inside this vacuum. \par 
Really, Higgs (vacuum) modes $ \Phi_a^{(n)}$, (\ref{Higgs-teta}), refer to the {\it rigid} axial symmetries group $G_1$ \cite{Al.S.}. \par
This group, in turn, compensates {\it local} rotations of the vacuum (YM-Higgs) configuration $(\Phi_a^{(n)},A_\mu^a)$ around the axis $z$. \par 
On the other hand,  this vacuum (YM-Higgs) configuration can be linked, in a continuous and smooth wise in its "Higgs" part, to Higgs vacuum BPS monopoles, disappearing at the origin of coordinates \cite{David3}.\par 
By that    two things are ensured. \par 
Firstly, the continuity (and smoothness) achieved now  for Higgs vacuum solutions, "legalizes"  collective solid  rotations inside  the Minkowskian  Higgs model with vacuum BPS monopoles quantized by Dirac \cite{Dir} (as a sign of thread topological defects inside  the  appropriate vacuum manifold $R_{\rm YM}$) within the $\epsilon$-neighbourhood of the origin of coordinates.  These rotations are described correctly by the action functional (\ref{rot}) \cite{David2}. \par 
Secondly, simultaneously, the Gauss law constraint (\ref{homo}) and the "colinearity condition" (\ref{colin})  are satisfied outward of this infinitesimal region with smooth vacuum "electric" monopoles (\ref {el.m}). \par 

\medskip
When we view very attentively Eq. (\ref{I}) \cite{David2} for the "rotary momentum" $I$ of the BPS monopole vacuum, we can note that the integration there is carried out over the infinite spatial volume $V\to\infty$. \par
 At neglecting the infinitesimally small interval $[0, \epsilon(\infty)]$ in this integral, one get nevertheless the same result. \par
  More exactly, the integral (\ref{I}) may be evaluated in the following way \cite{David2}:
 \be
  \label{kt}
I\simeq \frac{4\pi^2}{\alpha_s}\int\limits_{\epsilon}^{R}dr
~\frac{d}{dr}(r^2\frac{d}{dr}f_0^{BPS}(r)); \quad R\to \infty 
\ee.
It involves the "Higgs" BPS ansatz \cite{LP2,LP1}
\be\label{HBPS}  f_0^{BPS}(r)=\left[  \frac{1}{\epsilon\tanh(r/\epsilon)}-\frac{1}{r}\right],         \ee 
disappearing \cite{David3} at the origin of coordinates. 

In the paper \cite{David2} it was proposed the following approximation for $f_0^{BPS}(r))$:
\be \label{app} f_0^{BPS}(r)\sim 1-\frac \epsilon r \ee
with the asymptotics $ f_0^{BPS}(\epsilon)\to f_0^{BPS}(0)\to 0$, while $ f_0^{BPS}(\infty)\to 1$ (since $\epsilon$ also disappears in the infinite volume limit \cite{David2}). It should be taken $\epsilon(\infty)\to 0$ for $f_0^{BPS}(\epsilon)$ always in the above calculations. This result is very important and we will need it later.

We see that the extrapolation of Higgs vacuum BPS monopoles inward  the  interval $[0, \epsilon]$, where the potential superfluid nature of the  Minkowskian BPS monopole vacuum suffered the Dirac fundamental quantization \cite{Dir} is violated, does not affects the integrals (\ref{I}), (\ref{kt}) so long as  the integration limits in Eq. (\ref{kt}) are  $[\epsilon, \infty]$. \par 
This is just the region where the potential superfluid nature of that  vacuum takes place. \par 
It is determined, in the Dirac fundamental quantization scheme \cite{Dir}, by the Bogomol'nyi equation (\ref{Bog}), YM Gauss law constraint (\ref{homo}) and Gribov ambiguity equation (\ref{Gribov.eq}) as the  logical consequence \cite{rem2, LP2,LP1} of the Bogomol'nyi equation (\ref{Bog}). \par 
As to the interval $[0, \epsilon]$ (here, indeed, one must take $\epsilon=\epsilon(\infty)\to 0$),  integrating inside this interval comes to  integrating over the base of the  infinitely narrow cylinder along the axis $z$ (in the chosen rest reference frame), so long as the Higgs BPS ansatz 
$f_0^{BPS}(r)$ is a function of the  distance $r$ only. Inserting $f_0^{BPS}(r)$ in the approximation (\ref{app}) in (\ref{kt}) and changing interval ibid to $[0,\epsilon]$, we see that the  interval $[0, \epsilon]$  results a vestigial contribution to the  "rotary momentum" $I$, (\ref{I}), and one can neglect it actually.  \par
Indeed, due to required continuous stiching together Higgs vacuum BPS monopoles and Higgs "rotary" modes (\ref{Higgs-teta}) (and thus their covariant derivatives $D^2$), it is possible to write out explicitly the condition at which this  stiching together should occur.

As we have already mentioned in Introduction, this comes to evaluating the quantity $\delta$ entering Eq. (\ref{clas}) for $\phi(\rho)$.

Really, utilizing the look (\ref{HBPS}) for the ansatz $f_0^{BPS}(r)$, it is easy to get the following transcendental equation 
\be \label{sewing}
\frac 1{r^2}- \frac 1{\epsilon ^2 \sinh ^2 (\frac r \epsilon)} \sim r^{-2-2\delta}.
\ee
It is not a simple equation, and probably some numerical methods are needed for solving it. It is, however, beyond the scope of our work. 





\bigskip
On hand thus a delimitation between superfluid potential motions and collective solid rotations inside the  Minkowskian BPS monopole vacuum suffered the Dirac fundamental quantization \cite{Dir}. \par 
 Superfluid potential motions therein (controlled by the Bogomol'nyi and Gribov ambiguity equations) are referred actually to the spatial interval $r\in [\epsilon(\infty);
\infty]$, while collective solid rotations to the spatial interval $r\in [0, \epsilon (\infty)]$. \par
This delimitation is the next in turn evidence in favour of the first-order phase transition occurring in the Minkowskian  Higgs model with vacuum BPS monopoles quantized by Dirac \cite{Dir} and coming to the coexistence of two thermodynamic phases inside the  Minkowskian BPS monopole vacuum suffered the Dirac fundamental quantization. \par
The thermodynamic phases of superfluid potential motions and collective solid rotations are just such phases. \par
\medskip

To this stage of investigations about the Minkowskian  Higgs model with vacuum BPS monopoles quantized by Dirac, we cannot conclude regarding the concrete nature of the boundary between the mentioned thermodynamic phases as well as regarding various surface effects. \par
At the same time, we should like  note  that the transparent way to verify that the first-order phase transition occurs in any physical model is to compute the appropriate surface tension integral \cite{Landau52}
\be\label{surf}   \alpha_{12}=\int\limits_{-\infty } ^{\infty } (f_1-f_2)~ dx,\ee
where $f_1$ and $f_2$ are the free energies densities referring to two thermodynamic phases coexisting at the first-order phase transition occurring in the given physical model. \par 
\medskip 

\bigskip

The one more important lesson we have learned from our above discussion is that the Higgs field in here studied model  {\it loses its role as an order parameter} due to our claim that  "the covariant derivative $D\Phi$ of any vacuum Higgs field $ \Phi_a^{(n)}$ specified via (\ref{Higgs-teta}) merges with the covariant derivative of  vacuum Higgs BPS monopole solution". If we wish to construct (or if we presume) the gauge theory involving a first-order phase transition, such behavior of Higgs modes in no case the necessary behavior of order parameter, which must undergo a discontinuity.  And, on the contrary, the "magnetic" field squared, which has the intermittent shape 
\be \label{step1}
 <B>^2= \{ \aligned 
<B_1>^2&=0,\quad r\to 0;\\
 <B_2>^2&\neq 0 \quad r\to \infty \endaligned 
\ee
(where $r^2=x^2+y^2$), is a suitable "candidate" to be an order parameter for the  first-order phase transition anticipated in our YMH model with vacuum monopole solutions quantized by Dirac.

But it should be remembered that in the spatial region $r\neq 0$ ($r\to \infty$) the "magnetic" field ${\bf B}_1$ is completely determined by the Bogomol'nyi equation (\ref{Bog}), i.e. {\it again via the Higgs scalar} $\Phi$: more exactly, via its covariant derivative $D$ \footnote{Since $<B_2>^2$ is a gauge invariant, then also  $<\Phi>^2$ is gauge invariant.}.

Herewith the alone Higgs vacuum BPS monopole modes in the ${\bf x}\to \infty$  limit acquire  infinitely large masses limit and disappear from the spectrum
of physical excitations \cite{LP1}. This can be seen from the estimate \cite{LP1}
\be
\label{mass}
 \frac{1}{\epsilon}=\frac{gm}{\sqrt{\lambda}}=\frac{g^2<B_2^2>V}{4\pi}. \ee
for the natural and unique nonzero mass scale in the here discussed model. It is obvious from here that in the infinite volume limit $V\to\infty$ and when the coupling YM constant $g$ remains fixed (in the confinement region ${\bf x}\to \infty$).

Thus a specific transmutation occurs in the discussed model when the role of the order parameter is delegated from the Higgs VEV $<\Phi>^2$ to the VEV of the "magnetic" field squared $<B_2>^2$ determined by the Bogomol'nyi equation (\ref{Bog}): of course with its rotary "counterpart" $<B_1>^2=0$.

 This curious effect  is the next in turn "smile" of the "Cheshire Cat", Higgs vacuum BPS monopole modes: the alone Higgs vacuum BPS monopole modes disappear from the spectrum
of physical excitations at the spatial infinity, but its covariant derivative $D$ plays the important role in the discussed theory. 

As a striking proof of the said we can cite here the estimate for the "magnetic" field squared via the mass $\Delta m_\eta$ of the $\eta ^{'}$ meson given in the paper \cite{Pervush2} and based on the here discussed two phase gauge model with YM and Higgs BPS monopole and rotary modes:
 $$
 <B_2^2>=<D\Phi>^2=\frac{}{}\frac{2\pi^3F_{\pi}^2\triangle
 {m_\eta}^2}{N_f^2\alpha_s^2}=\frac{0.06 GeV^4}{\alpha_s^2}~.
 $$
In this equation $F_{\pi}\approx 130.41 {\rm Mev}$ is the pionic decay constant and $N_f$ is the number of flavours. We send our readers  for details of this calculation to Section 5.10 of the paper \cite{Pervush2}.

The reasonable question may be asked in this context. Does it happen the above discussed transmutation in another gauge models involving Higgs and YM (vacuum) BPS monopole modes (i.e. those satisfying the Bogomol'nyi equation)?. A good example of such models is the case of the $SU(5)$ spontaneous break-down in GUT to the product $SU(3)\otimes SU(2) \otimes U(1)$ of symmetries group which occurs at the temperatures in the interval $10^{14}- 10^{11}$ Gev. On the author opinion, such a high temperature scale is an important factor in answering correctly this question.

\bigskip
The important consequence of the  presence of topologically nontrivial threads in the Minkowskian Higgs model with vacuum BPS monopoles quantized by Dirac \cite{Dir} is annihilating \cite{Al.S.} two equal magnetic charges $  {\bf m}_1  =  {\bf m}_2={\bf m}(n)\neq 0$ ($n\in{\bf Z}$) colliding at crossing a rectilinear topologically nontrivial thread $A_\theta (n)$: (\ref{Ateta}), (\ref {Aro}). \par 
This effect is the particular case of changes in the relative sign  of two magnetic charges as the integration region $\Gamma$ in (\ref{mul}) intersects a (rectilinear) thread. \par
Suppose that the magnetic charge $\vert {\bf m}\vert =m_1$ is concentrated to the left of the thread $A_\theta (n)$, while the magnetic charge $\vert {\bf m}\vert =m_2$ is concentrated to the right of this thread. Herewith we  concentrate these magnetic charges in points. \par 
Concentrating magnetic charges in points does not change the matter, despite  magnetic charges are distributed volume actually due to (\ref{mul}). \par
Due to (\ref {dvaznaka}), the both  magnetic charges are specified to within their signs.
\par
One can  draw a closed curve  connected the both magnetic charges (see Fig. 15 in \cite{Al.S.}). It consists of  two lines, $\gamma_1$ and $\gamma_2$, between the charges,  one rounds in the opposite directions. \par 
This implies that the vacuum "magnetic" tension $\bf B$ (specifying via the Bogomol'nyi equation (\ref{Bog}))  can change its sign depending on the "direction": either $\vec \gamma_1$ or $\vec \gamma_2$, in which one crosses the thread. \par 
Really, we should always consider the relative sign of the charges ${\bf m}_1$ and ${\bf m}_2$: more precisely, our interest is the difference $m_1-m_2$ as we move in the   "direction"  $\vec \gamma_1$; contrariwise, our interest is the difference $m_2-m_1$ as we move in the  opposite  "direction"  $\vec \gamma_2$.\par 
Taking account of the duality (\ref {dvaznaka}) 
in the definition of a magnetic charge, we encounter few "alternatives" for the values of the relative magnetic charge in both the cases. \par
The most interesting case is when $m_1=m_2$. \par
In this case  some two vacuum BPS monopoles belonging to the one fixed topological class $n\in \bf Z$ annihilated mutually at approaching each  other crossing the topologically nontrivial thread $A_\theta $ lying in the Minkowskian YM vacuum manifold $R_{YM}$, (\ref{RYM})(the same is correct also for excitations over the BPS monopole vacuum). \par 
The above reasoning demonstrates   that magnetic charges  are not preserved  in the Minkowskian Higgs model with vacuum BPS monopoles quantized by Dirac. \par 
Magnetic charges cease then to be motion integrals;  at the same time they are still topological invariants\rm: continuous deformations do not   change the difference $m_2-m_1$. \par
\medskip 
In the monograph \cite{Al.S.}, in \S $\Phi13$, there was proposed the following, more general,  definition of a magnetic charge, taking account of rectilinear (topologically nontrivial) threads in appropriate non-Abelian gauge models. \par
 One considers a membrane pulling  on the rectilinear thread $A_\theta $ lying along the axis $z$ of the chosen coordinate system.  Then  one removes this membrane from the Minkowskian space.  This results a one-connected set, on which one can choice a fixed continuous branch of the magnetic tension $\bf B$. \par
In fact, this "trick" comes to the restriction  of the integration region in Eq.  (\ref{mul}) for a magnetic charge by the "left" (respectively, "right") semi-spaces of the Minkowskian space with respect to the axis $z$. This ensures that magnetic charges are  preserved  in each of these semi-spaces taken separately. \par
\medskip
In the present study we  don't have for our object deriving the concrete contribution from the above discussed annihilating effect \cite{Al.S.} for equal topologically nontrivial magnetic charges colliding at crossing a thread $A_\theta $ in the total action functional (Hamiltonian) of the quested Minkowskian Higgs model with vacuum BPS monopoles, although such contribution exists undoubtedly. \par
Nevertheless, now we should like make some remarks which may be helpful for finding such    contribution. \par
For instance, in the paper  \cite{Linde1} possible interactions between (YM) magnetic monopoles and antimonopoles were studied involving forming 
so-called {\it monopole molecules}.\par 
Herewith only two types of monopole molecules are possible \cite{Linde1}, reflecting ordinary Feynman rules  for gauge non-Abelian fields (see e.g. \S 9.1 in \cite{Cheng}).\par 
There are, firstly, confined colourless magnetic monopoles $M$ and $\tilde M$ (with the symbol $M$ referring to a  monopole with the magnetic charge $\bf m$ and the symbol $\tilde M$ referring to an antimonopole with the magnetic charge $-\bf m$, respectively).  These monopole molecules were also referred to as monopole-antimonopole  {\it mesons} in Ref. \cite{Linde1}. 
The well-known mechanism for the monopole-antimonopole annihilation is via  Abrikosov threads  of the "magnetic" field \cite{Al.S.,Ryder, Abrikosov} (see also \S 6.2 in \cite{Linde}).\par
Secondly \cite{ Linde1, Cheng}, confined colour triplet and antitriplet of monopoles are also possible in the nature. \par
Herewith $MMM$ (respectively, $\tilde M \tilde M \tilde M$) states were referred to as {\it baryons} in \cite{ Linde1}.\par
The presence of topologically nontrivial threads in a gauge non-Abelian theory (for instance, rectilinear threads (\ref {Ateta}) \cite{Al.S.}), implying the annihilation of monopole pairs with {\it equal} magnetic charges colliding at crossing such a (rectilinear)  topologically nontrivial thread, changes substantially the matter in comparison with the case of the Abrikosov mechanism \cite{Ryder,Linde, Linde1} of the monopole-antimonopole confinement. \par
The former case is rather similar to the case of confined colour triplet and antitriplet of monopoles (where the role of the one field in such triplets is played by the rectilinear thread $A_\theta$). \par
On the other hand, the computations about the annihilation of monopole pairs with { equal} magnetic charges (at least at the level of Feynman diagrams)  generalizes somewhat the computations \cite{BLP,A.I.} about electron-positron pairs in QED. \par
Herewith unlike QED, where the creation-annihilation processes are forbidden with an only one photon taking part in such processes (by the reason of maintaining the four-momentum in a Feynman diagram involving two fermions and one photon on-shell: see e.g. \S 25.1 in \cite{A.I.}), in an YM model involving only massless gauge fields, processes with three YM fields are quite permissible. \par
In particular, it is correctly for the Minkowskian Higgs model with vacuum BPS monopoles quantized by Dirac, where YM fields, transverse topological Dirac variables  $A_i^D$ ($i=1,2$), are specified \cite{rem2} over the light cone $p^2=0$ due to the removal  \cite{David2} of temporal YM components. \par 
\bigskip
The just discussed annihilation mechanism \cite{Al.S.} for (equal) magnetic charges in the Minkowskian Higgs model with vacuum BPS monopoles quantized by Dirac \cite{Dir} can cause disappearing, in a definite  time interval $\tau$, topologically nontrivial magnetic charges \linebreak ${\bf m}\neq 0$.\par 
Let us suppose that this has occurred indeed. \par 
In  this case, as it is well known, the Dirac quantization \cite{ Dirac, Ryder, Cheng} of magnetic and electric charges, taking the  general look \cite{Hooft} 
\be \label{Dir.con}
e_1{\bf m}_2- e_2{\bf m}_1 = \frac{1}{2}n; \quad n\in {\bf Z};
\ee
for two interacting (BPS, 't Hooft-Polyakov and so on) monopoles in a  Minkowskian Higgs model, implies for Higgs  monopoles to have arbitrary electric charges $e$ associated with the topological number $n=0$ (and thus with magnetic charges ${\bf m}=0$). \par 
In the review \cite{Hooft} (and in other modern physical literature) such situation in a $SU(n)$ non-Abelian gauge theory (involving the spontaneous breakdown of the initial $SU(n)$ gauge symmetry down to its $U(1)^{n-1}$ subgroup)  when gauge fields $A$ possess zero magnetic charges, that implies  arbitrary electric charges for Higgs modes $\Phi$ due to the Dirac quantization \cite{Dirac} of  electric and magnetic charges, was referred to as the so-called \it Higgs phase\rm.  \par  
In the notion "the Higgs phase", one implies (for instance, in Ref. \cite{Hooft}) however  that  all magnetic charges are confined by (infinitely narrow) Meissner flux tubes, similar to ones in a  superconductor \cite{Landau52, Cheng}. \par  
  In  turn, this involves the linearly  increasing "Mandelstam" potential $O(Kr)$ (with $K$ being the string tension) between  YM monopole and antimonopole. \par  
As to Higgs vacuum BPS monopole solutions, they carry free electric charges in the Higgs phase \cite{Hooft}. \par
Herewith in the concrete case of the Minkowskian Higgs model with vacuum BPS monopoles quantized by Dirac, the topologically trivial Higgs BPS monopole modes contribute to the "electric" energy (\ref{rot}) \cite{David2} of the appropriate BPS monopole vacuum, induced by its topological rotations. \par 
\medskip
There is an interesting and somewhat curious occurs with  topologically trivial Higgs vacuum BPS monopoles in the Higgs phase. \par 
On the one hand, they induce vacuum "electric" monopoles (\ref{el.m}). Herewith the term "electric" is highly relative in  this context since this term is associated with  temporal vacuum YM components $Z_a$ \cite{Pervush1}, (\ref {zero}), solutions to the Gauss law constraint (\ref{homo}), having a definite (rather mathematical) analogy with temporal  components, $F_{0\mu}$,  of Maxwell tensors, i.e. with  electric strengths $\bf E$. \par 
On the other hand, in the Higgs phase \cite{Hooft}, topologically trivial Higgs vacuum BPS monopoles $\Phi_{(0)}^a$ carry free electric charges, that are told immediately to vacuum "electric" monopoles (\ref{el.m}). \par 
Thus vacuum "electric" monopoles (\ref{el.m}) become indeed \it electric \rm fields in the Higgs phase.
\par 
\medskip
It should be distinguished indeed between the conception of the Higgs phase understood customary (e.g. in Ref. \cite{Hooft}) and that in the Minkowskian Higgs model with vacuum BPS monopoles quantized by Dirac \cite{Dir}.\par
First of all, the conception of the Higgs phase, as it is understood in \cite{Hooft}, is based on  fixing {\it the maximum Abelian gauge} (MAG) and on final violating (in the simple YM case) the initial $SU(2)$ gauge symmetry in the
\be \label{NO}
SU(2)\to U(1)\to {\bf Z} 
\ee 
wise, involving the "discrete multiplier" $\bf Z$.\par
Now let us discuss briefly both these features of the Higgs phase (as it is understood in modern physics). \par 
Fixing MAG comes to singling out the maximum Abelian ($\sim U(1)$) subgroup in $SU(2)$. \par 
The simple way to do this is associated with the diagonal Pauli matrix $\tau_3$.\par 
Then, $\tau_3$ would be considered as the generator of the maximum Abelian subgroup in $SU(2)$. \par 
This implies maximizing the integral
\be \label{int1} \int d^4x \left [(A_\mu^1)^2+  (A_\mu^2)^2\right ]; \quad A_\mu = A_\mu^a \tau_a, \ee
or, that is equivalent (see e.g. Ref. \cite{Jeff})
of maximization of the quantity
\be \label{Jeff2} 
      \sum_x \sum_{\mu=1}^4 \mbox{Tr}[\tau_3 U_\mu(x) \tau_3 U^\dagger_\mu(x)].
\ee
On the other hand, the said is equivalent to decomposing an YM field into off-diagonal and  diagonal components \cite{Sobr1}:
$$ {\cal A}_\mu= A_\mu^a \tau_a +  A_\mu \tau_3 ; \quad a=1,2.     $$
Similarly, one find 
$$  {\cal F}_{\mu \nu} = F_ {\mu \nu}^a  \tau_a  + F_ {\mu \nu}\tau_3 $$
for the field (YM) strength 
with off-diagonal and diagonal parts given by 
$$  F_ {\mu \nu}^a  = D_\mu^{ab} A_\nu^b-  D_\nu^{ab} A_\mu^b  $$
and 
$$  F_ {\mu \nu}  = \partial_\mu A_\nu-  \partial_\nu A_\mu,        $$
respectively.

For the YM action (in the Euclidean space $E_4$) one get finally 
$$ S_{\rm YM}= \int d^4x (F_ {\mu \nu}^a F_ {\mu \nu}^a)+  F_ {\mu \nu}  F^ {\mu \nu}.            $$
MAG may be got \cite{Sobr1} by imposing the Lorentz gauge
$$  D_\mu^{ab} A_\mu^b  =0.    $$
\medskip
MAG in YM theories can be generalized easy for the case of the initial $SU(3)_{\rm col}$ gauge symmetry group in "realistic" QCD models, violated then in the
\be \label{QCD} SU(3)_{\rm col}\to U(1)\otimes  U(1) \ee 
wise.

In  this case MAG can be fixed by singling out the Gell-Mann matrices $\lambda_3$ and $\lambda_8$ from the total set of the Gell-Mann matrices $\bf \lambda$.

By analogy with (\ref{int1}), now MAG fixing is equivalent to maximizing the integral
\be \label{int2} \int d^4x \left [(A_\mu^1)^2+  (A_\mu^2)^2+  (A_\mu^4)^2+  (A_\mu^5)^2+   (A_\mu^6)^2+  (A_\mu^7)^2 \right ]; \quad A_\mu = A_\mu^a \lambda_a.                                                  \ee
\medskip 
However to achieve the quark confinement in the YM theory and especially in "realistic" QCD models, involving the initial $SU(3)_{\rm col}$ gauge symmetry group (violated then in the (\ref{QCD}) wise), the appropriate Abelian $U(1)^{n-1}$ ($n=2,3$) subgroups would be also {\it broken down further}.

For  the initial $SU(2)$ gauge symmetry group this occurs in the (\ref{NO}) wise, while for the initial $SU(3)_{\rm col}$ gauge symmetry group, it is
\be \label{NO1}
 SU(3)_{\rm col}\to U(1)\otimes  U(1) \to {\bf Z}.
\ee
The said can be grounded by the reasoning to provide the needs of  the quark confinement in the above theories.

The sure sign of  the quark confinement taking place is, at is well known, the {\it area law} \cite{Cheng} satisfied by {\it Wilson loops}, taking the typical shape
\be \label{area} W(C)\sim\exp\{ -KA(C)  \}, \ee
implicating the areas $A(C)$ of surfaces enclosed by appropriate paths $C$.

Just in order implementing the area law (\ref {area}), characterizing the confinement of quarks and gluons, the phenomenological picture in which quark $q$ ($\bar q$) and gluons $A$ are confined by (infinitely thin) Meissner flux tubes is quite fit.

This phenomenological picture is similar to that proper to superconductors.

In particular, in superconductors, the magnetic field $\bf B$  decays exponentially at penetrating into the superconducting region \cite{Lenz}: 
\begin{equation}
   \label{lon}
{\bf B}(x)= {\bf B}_{0}e^{-x/\lambda_{L}},
 \end{equation}  
with
\begin{equation}
   \label{pede}
\lambda_{L} = \frac{1}{M_{\gamma}}   
 \end{equation}
being the {\it penetration or
London depth}, depending on the "photon mass"
\be \label{photomass}
M_\gamma=\sqrt 2~ ea.
\ee
The parameter $a$ appears in latter Eq. 

It plays the role of the order  parameter in the {\it Ginzburg-Landau} superconductivity model \cite{Ryder,Lenz,V.L.}: $a=0$ for the normal phase, while $a\neq 0$  for the superconducting phase. \par
The magnetic field $\bf B$, (\ref{lon}), decaying  exponentially at penetrating into the superconducting region, is the solution to the {\it Maxwell-London equation} \cite{Ryder,Lenz} 
\be \label{London}  \mbox{rot}\,{\bf B}=\mbox{ rot}\,\mbox{rot}\,{\bf A} = {\bf j}_s= 2 e^2 a^2 {\bf A},\ee
implicating the preserved Noether current \cite{Landau52, Ryder} 
\be \label{Noether current}
{\bf j}_s=- i\frac{e}{ m} (\phi^* \nabla \phi- \phi\nabla \phi^*)- \frac{e^2}{m} \vert\phi  \vert ^2 {\bf A},
\ee
with $\phi $ being the superconductor wave function. \par
Eq. (\ref{lon}) serves as the mathematical expression for the Meissner effect. The essence of this effect is in expulsion of the magnetic field from the superconducting region. \par
\medskip
Similar locating the "magnetic" flux in the shape of infinitely narrow tubes explains the confinement effect in QCD. \par 
The {\it confinement phase} can be treated \cite{Hooft} as dual to the Higgs phase, discussed above. \par 
In this phase gluons and quarks, that are purely electric objects, are confined by infinitely narrow "magnetic"  flux tubes (strings). Herewith Higgs fields become purely magnetic objects. \par 

The occurred in the confinement phase requires undoubtedly the spontaneous breakdown of the initial ($SU(3)_{\rm col}$ or $SU(2)$) gauge symmetries groups with the appearance of Higgs modes. \par 
In this case also MAG fixing seems to be quite natural. \par 
In the modern literature such (phenomenological) picture of confinement based on MAG fixing and  the  {\it dual-superconductor
picture} (similar to that one can observe in the Ginzburg-Landau model \cite{V.L.} and implicating \cite{Hooft} the Higgs and confinement phases) is referred to as the {\it Abelian dominance} \cite{Suzuki}). \par
\medskip
But, indeed, it is not sufficient to violate the initial $SU(n)$ ($n=2,3$) gauge symmetry group in a non-Abelian model up to its $U(1)^{n-1}$ subgroup in order to ensure the quark confinement in that model. \par
The  thing is that the appearance of (infinitely thin) "magnetic"  flux tubes (strings) confining quarks in QCD can be explained good in the framework of the well-known Nielsen-Olesen model \cite{No}.\par
This model serves as a highly correct  representation for the Abelian  Higgs theory. \par 
With the example of the initial $U(1)$ gauge symmetry group afterwards violated, in the original paper \cite{No} it was demonstrated the existence of specific solutions to the equations of motions, {\it Nielsen-Olesen vortices} (see e.g. \cite{Ryder,Lenz}). \par
The way grounding the existence of Nielsen-Olesen (NO) vortices in the Abelian  Higgs theory comes to violating the initial $U(1)$ gauge symmetry group proper to that theory (this involves the nonzero value $\vert \phi\vert \neq 0$ for the Higgs field $\phi$ in the "asymmetric" phase, entering explicitly \cite{Ryder,Lenz} the appropriate equations of motions) and, from the topological standpoint, to the presence of thread topological defects (just interpreted as NO vortices) inside the appropriate vacuum manifold. \par
\medskip
In the light of the said, the task to ascertain the look of   this vacuum manifold becomes very important. \par
The investigations about the NO model \cite{No} are not the immediate goal of the present study, but some outlines and conjectures concerning  this model will be helpful for us in understanding of processes taking place in QCD if stand  on the Abelian dominance viewpoint \cite{Suzuki}. \par 
\medskip
From our previous discussion about the liquid helium II theory, some analogies suggest concerning the origin of NO vortices in the  Abelian Higgs model \cite{No}.\par
First of all, from the topological point of view, NO vortices, as a particular case of  thread topological defects, would be associated with the "discrete" vacuum geometry should be assumed for the appropriate degeneration space. \par
Such degeneration space (vacuum manifold) can be obtained in the way similar to that in which the discrete vacuum manifold $\tilde R$ has been got  in the rested liquid helium II specimen case. \par
However there are some principal distinctions between the latter case and the  Abelian Higgs model \cite{No} involving NO vortices. \par
So, for instance, the first-order phase transition occurs primary in that model (see the arguments \cite{Linde}, repeated also in Ref. \cite{ rem1}). It comes to a discontinuity in the plot of the appropriate order parameter (as a such order parameter, the vacuum expectation value $<\phi^2>$ for the Higgs field 
$\phi $ can be chosen). This implies the coexistence of the metastable (symmetric) and stable (asymmetric) thermodynamic phases in the  Abelian Higgs model \cite{No}.
It is just the supercooling case (according to the argument similar to those \cite{Nels} we have encountered above at the analysis of vortices in a liquid helium II specimen). \par
Once thus, the way  (\ref{tilde}) violating the $U(1)$ gauge symmetry (fit for the {\it second-order phase transition} \cite{Levich1} in a helium  specimen)  is not quite suitable in the  Abelian Higgs model \cite{No} with  NO vortices implying the {\it first-order phase transition}. \par
In the first case, the  symmetric thermodynamic phase disappears entirely upon $U(1)$ violating. \par 
This makes quite correct Eq. (\ref{tilde}) for the quite separate set \cite{Engel} $\tilde U$, coinciding with the degeneration space (vacuum manifold) $ R_r$.\par 
The look of this discrete space (in which all the topological domains refer to the asymmetric thermodynamic phase) just represents the above (primary) second-order phase transition in the  helium theory. \par
\medskip 
Unlike the former case, in the  Abelian Higgs model \cite{No}, in  which the { first-order phase transition} occurs involving coexisting \cite{ rem1,Linde} the stable and metastable thermodynamic phases (at the fixed temperature $T$), it may be assumed  gradual {\it thermal} destroying the $U(1)\cong S^1$ gauge symmetry in the following way. 

\medskip Note firstly that coexisting two thermodynamical phases course the first-order phase transition required means that we should retain the $U(1)$ gauge symmetry and simultaneously forming the stable asymmetric  thermodynamic phase represented by a discrete degeneration space of the $\tilde U$ wise. It turns out that this task is quite solvable! 

We must remember that the gauge $U(1)$ group space is isomorphic to the circle $S^1$. In any topological sector of this circle (say, in $n^{\rm -th}$), it is represented, in the "loop framework" \cite{Postn4}, by the appropriate homotopical class $[u]^n$, (\ref{cll}). Let's speculate what will happen at deleting some loops in several (or even each!) homotopical classes $[u]^n$. Just this promotes forming the stable asymmetric  thermodynamic phase ala $\tilde U$. The topology, i.e. the gauge symmetry, is maintained when we remove some  loops $\gamma^n \in [u]^n$. Herewith we simply restrict the set of available representatives in the class $[u]^n$, the topology of $S^1\cong U(1)$ {\it does not change}! And this is a desired result.

We are also interested in the case when  the gauge $U(1)$ group space is distorted (with forming NO vortices). Let us discuss now   under what circumstances this can occur. This can occur when

 $\bullet$  a single point is removed. Then $S^1/\{p\}\cong {\bf R}$ is contractible. Fundamental group becomes trivial.

$\bullet$ when removing countably many points: still disconnected, possibly totally disconnected depending on density.

$\bullet$ at removing  dense subsets: we may destroy path-connectedness or compactness.

\medskip Indeed, this "coexistence" of  the metastable and stable thermodynamic phases in the  Abelian Higgs model \cite{No} with NO vortices cannot last infinitely long time. \par
After all (generally, at lowering the temperature below the Curie point $T_c$), the Higgs field $\phi$ in the  Abelian model \cite{No}, leaving the "false" vacuum  $\phi=0$,  rolls into its "true" minimum (say, $\phi_0\neq 0$),  and the metastable (symmetric) thermodynamic phase (corresponding to the above described continuous set and the "false" vacuum $\phi =0$) disappears herewith entirely. \par
From the "geometrical" standpoint, the gradual destruction of the  continuous \linebreak $U(1)\simeq S^1$ group space (with forming the "bubble" \cite{Linde} of the new, stable and "discrete", thermodynamic phase inside the metastable, "continuous", one) occurs. \par
At this disappearing the metastable (symmetric) thermodynamic phase, the accumulated latent heat is released; it is just the reheating situation \cite {Coles}.\par
It may be guessed that rolling the  Higgs field $ \phi$ in the "true"  minimum $\phi_0$ corresponds to the dual-superconducting picture of confinement in QCD based upon MAG fixing and the Abelian Higgs model \cite{No}  with NO vortices. 
 
\bigskip 
In connection with our discussion  about the   Abelian Higgs model \cite{No} (implicating NO vortices), it will be relevant to mention the very important analogy (in  the sphere of the topology and phenomenology) between this model and the case \cite{Landau52,Lenz,V.L.} of  superconductors. \par 
For instance, in Type II superconductors, the second-order phase transition occurs with definite conditions in the presence of an background magnetic field $H$. \par 
First of all, it would be therein  \cite{Landau52,Lenz,V.L.} $\kappa >1/\sqrt 2$ for the Ginzburg-Landau parameter 
$$ \kappa =\frac{\lambda_L}{\xi}, $$ 
with $\lambda_L $ given in (\ref{pede})  \cite{Lenz} and $\xi$ being the coherence length which physical sense is \cite{Landau52} the correlations radius of fluctuations in the order parameter $<\phi^2>$.\par 
Generally speaking \cite{Landau52}, the Ginzburg-Landau parameter $\kappa$ is a function of the temperature $T$:  $\kappa\equiv\kappa (T)$. 
As it was demonstrated in the original paper \cite{V.L.} (see also \cite{Landau52}), the inequality  $\kappa >1/\sqrt 2$, true for Type II superconductors, is associated with the negative surface  tension $\alpha_{ns}<0$  (which look is (\ref{surf})) between the normal (n) and superconductor (s) phases. 

In the latter case it becomes advantageous, from the thermodynamic standpoint, to compensate increasing the volume energy by  this negative surface  energy $\alpha_{ns}<0$.
This implies arising germs of the n-phase inside the s one at the values $H$ of the background magnetic field $H$ exceeding a field  $ H_{c_1}$, called \cite{Landau52} {\it the   lower critical field}.
Vice verse,  germs of the s-phase inside the n one arise at the values $H$ don't exceeding a field  $ H_{c_2}$, called {\it the   upper critical field}.

Thus in the interval $ H_{c_1}<H< H_{c_2}$, one can observe \cite{Landau52} the {\it mixed state} in Type II superconductors (there are definite alloyed metals and combinations of metals). In this case a Type II superconductor is simultaneously in the n and s states.
At $H\leq H_{c_1}$, a specimen is purely in the s-state, while it is purely in the n-state at $H\geq H_{c_2}$.

\bigskip
It turns out that the {\it second-order phase transition} takes place in this situation, and now we would like to clarify this.

First of all, it can be argued (see, for example, the work \cite{Kaj})  that in type II superconductors vortex lines of the n-phase {\it repel each other}; thus the vorteces tend to form a vortex lattice in
the type II region. In fact, due to the periodic lattice boundary conditions, even one vortex forms a
square lattice with its periodic counterparts.

The main argument in favour of one or another kind of phase transitions occurring in a medium is the analysis of second derivatives of thermodynamical potentials by the temperature and pressure (for instance, the heat capacity and compressibility), which change abruptly if the  second-order phase transition takes place. In the same time, the first derivatives (such as the energy and volume of the medium) do not change.

The second case, of the free energy plot, has been analyzed, with the example of the type II superconductor, in the monograph \cite{tinkman}. 
In this monograph the behavior of the magnetization curve $B(H)$ was analyzed in the vicinities of the critical points $H_{c1}$ and  $H_{c2}$. This is in fact the same as the analysis of the Gibbs free energy $G$ in the external magnetic field $H$ switched-on, which contains always the item $(BH)/4\pi$.

Such behavior of the magnetization curve $B(H)$ is closely related to the  picture of vortices in the studied superconductor. 

\medskip So, three regimes \cite{tinkman} can be distinguished in the $[H_{c1}, H_{c2}]$ interval.

1. Very near  $H_{c1}$, $\Phi_0/B\gg \lambda_L$ for the elementary flux $\Phi_0=\pi \hbar c/\vert e \vert$. Then the vortices are separated by distances more than $\lambda_L$. In this case only a few neighbors are important.

2. For moderate values of $B$, such that $\xi^2 \ll \Phi_0/B \ll  \lambda_L^2$, many vortices appear within interaction range of any given one; this generate the $\sum _{i>j} F_{ij}$ contribution into the Gibbs energy $G$ \footnote{Following \cite{tinkman}, cite the explicit look for $F_{ij}$. This is 
$$F_{ij}\sim \frac{\Phi_0^2}{8\pi^2\lambda_L ^2} \ln ( \frac{\lambda_L}{\xi})$$ It is positive when $\lambda_L > \xi$. But we know already that the type II superconductivity begins when  $\kappa=\lambda_L/\xi >1/\sqrt 2$ ($1/\sqrt 2\approx 0.714$). Herewith the interval $\kappa \in [1/\sqrt 2, 1]$ represents a {\it transitional regime} between type-I and type-II superconductivity. In this interval, the surface energy is small and negative, meaning vortex formation is possible but energetically delicate. Vortices may form, but they are not strongly repulsive.
The interaction between vortices can be weakly attractive or neutral, depending on the exact value of  $\kappa$. This can lead to clustering or nonuniform vortex arrangements, unlike the regular Abrikosov lattice seen in higher $\kappa$ type-II superconductors.
}. However, it is still a good approximation to neglect details of the core. 

3. Near  $H_{c2}$, $\xi^2\approx \Phi_0/B$, so that the cores are almost overlaping.

\medskip Just at the careful analysis of the "regime one" one can make sure that the second order phase transition occurs in an (infinitesimal) neighbourhood of $H_{c1}$. The (long enough) calculations \cite{tinkman} give the following look of the magnetic field $B$ in a neighbourhood of $H_{c1}$:

\be \label{B1}
B=\frac{2\Phi_0}{\sqrt 3  \lambda_L^2} \{ \ln [\frac{3\Phi_0}{4\pi  \lambda_L^2(H-H_{c1})}] \}^{-2}.
\ee
$B$ is {\it continuous} at $H_{c1}$, corresponding to a {\it second order phase transition}  (since $B$ enters the expresiion for the free energy expression of the model studied via the $(BH)/4\pi$  item).

The similar arguments in favor of the  second order phase transition occurring are applied to the regime near $H_{c2}$. More exactly, in \cite{tinkman}
the magnetization curves $4\pi M=f(H)$ for different values of $\kappa$ were considered (see Fig. 5.2. in \cite{tinkman}). This analysis of curves shows that only at $\kappa >1/\sqrt 2$ a  second order phase transition takes place.

\medskip The attraction of vorteces in type one superconductors leads \cite{Kaj} to forming the cylindrical hole (the "broken" s-phase) inside the
magnetic field configuration (the symmetric n-phase). 

For this case, the discontinuity $\Delta \frac{\partial G/V}{\partial y}$ of the canonical free energy $G(x,y,H/e_3^3)$, with \cite{Kaj} $y$ and $x$ being two dimensionless ratios

\be \label{xy}
y=\frac{m_3^2(e_3^2)}{e_3^4}, \quad x=\frac{\lambda_3}{e_3^2}
\ee
(involving the electron mass $m_3$ and the Cooper pair interaction constant $\lambda_3$); $e_3$ is a scale introducing in the model by the authors \cite{Kaj}.

Indeed, 
\be \label{dege}
\Delta \frac{\partial G/V}{\partial y}=e_3^4 \Delta <\phi^\star\phi>; \quad \Delta\frac{\partial G/V}{\partial H/e_3^3}=-e_3^3 \Delta B,
\ee
with $\phi$ being the macroscopic instantaneous wave function associated with the  Cooper pair.

For fixed $x$ the latent heat $L$ of the transition is defined as the discontinuity in the "energy" variable $E$ obtained from $G$ by a Legendre transformation with respect to $y$, $H/e_3^3$:
\be \label{latent}
L=\Delta E= -y \Delta \frac{\partial G}{\partial y} -H_c \Delta \frac{\partial G}{\partial H}= V[-y e_3^4 \Delta <\phi^\star\phi> +H_c \Delta B].
\ee

For fixed x, the identity $\Delta G(x,y, H_C)=0$ leads to the Clausius-Clapeyron equation, relating the different discontinuities:
\be \label{Claus}
\Delta \frac{\partial G}{\partial y}=-\frac{\partial H_c}{\partial y}\frac{\partial G}{\partial H}\Leftrightarrow e_3^4 \Delta <\phi^\star\phi>=\frac{\partial H_c}{\partial y} \Delta B.
\ee
{\it Thus it is enough to measure one of the discontinuities, and the curve} $H_c(y)$. 

\bigskip   From the topological (geometrical) point of view, the said suggests two ways for breakdown the  $U(1)$ gauge symmetry inherent to the superconductivity model. 

In Type II superconductors, in which the second-order phase transition takes place, the appropriate $U(1)\simeq S^1$ group manifold collapses {\it instantly} in the way when domain walls arise simultaneously between all the topological sectors of the circle $S^1$.

In Type I superconductors, in which the first-order phase transition takes place, collapsing the $U(1)\simeq S^1$ group manifold occurs {\it gradually} in the way similar that in the case of NO vortices, us discussed above.  A distinctive feature of the  first-order phase transition taking place in Type I superconductors thus the coexistence of the mentioned geometrical (topological) structures (at the temperature $T\to 0$). Note herewith that the latent heat depends manifestly on the change in the background magnetic field $H$ (rather on changes in the temperature of environment!) as Eq. (\ref{latent}) reads. Thus we can speak about the gradual "magnetic" distortion of the $U(1)\cong S^1$ group space in this case. This is instead of the thermal distortion in the case of the NO model \cite{No}.


\bigskip In spite of the said, it is obvious that in the Curie point $T_c$, $ H_{c_1}\to H_{c_2}\to 0$. The said allows, for all that, to construct the plots $H(T)$ for the s, n and mixed states in Type II superconductors (see Fig. 7 in \cite{Landau52}).   

\medskip
It is enough manifest that the case of Type II superconductors may be treated as a nonrelativistic limit of the Abelian Higgs NO model \cite{No} \footnote{It is so at identifying \cite{Ryder, Lenz} the density of  superconducting {a \it Cooper pair} with the Higgs field squared $\vert \phi\vert $ in the NO model \cite{No}. }.



Indeed, it is a similarity  between the helium at rest theory \cite{Halatnikov} and the Type II superconductivity \cite{Lenz,V.L.}, namely  that in both the cases the $U(1)\simeq S^1$ group space is  destroyed instantly entirely (in the (\ref {tilde}) wise)
down to the discrete (quite separate) set $\tilde U$ involving domain walls between topologies. 


\medskip
The outlined above parallel between the Type II superconductivity and the Abelian Higgs NO model \cite{No} promotes also the comprehension of the  
dual-superconductor
picture \cite{Hooft,Lenz} of the confinement of quarks and monopoles. 

One can imagine \cite{Lenz}, in his thoughts, placing some north and
south magnetic monopole inside a type II superconductor in such a wise that they are separated by the distance $d$.
Thinking that the magnetic
field is concentrated in  cores of  vortices and will not extend into the superconducting region, the field energy of this system becomes \cite{Lenz}
\begin{equation}
  \label{Vmm}
  V= \frac{1}{2}\int d^3 x\, {\bf B}^2 \propto \frac{4\pi d}{e^2 \lambda_L^2}. 
\end{equation}
Thus, the interaction energy of  magnetic monopoles grows linearly with their separation, as it would be in the  
dual-superconductor
picture \cite{Hooft} of confinement. 

\bigskip
The Minkowskian Higgs model with vacuum BPS monopoles quantized by Dirac \cite{Dir} results the approach to QCD (including the confinement picture) rather another that one gets fixing MAG and assuming the  
dual-superconductor
picture \cite{Hooft} of confinement (implicating the Mandelstam linearly increasing potential). 

Firstly, as it was discussed in Refs. \cite{ David2, Pervush2} (see also \cite{ fund}), at constructing Minkowskian Gauss-shell QCD (generalizing the Minkowskian Higgs model with vacuum BPS monopoles quantized by Dirac and involving the initial $SU(3)_{\rm col}$ gauge symmetry group), the specific gauge for the "intermediate" $SU(2)_{\rm col}$ symmetry group in the "breakdown chain"
$$  SU(3)_{\rm col}\to  SU(2)_{\rm col} \to U(1)     $$
can be fixed.

It comes \cite{ fund, David2, Pervush2} to the choice of the antisymmetric Gell-Mann matrices $\lambda_2$, $\lambda_5$, $\lambda_7$ as generators for $SU(2)_{\rm col}$.

At further assuming the  discrete geometry (\ref{fact2}) for $ SU(2)_{\rm col}$, this involves lot of consequences for Gauss-shell QCD (described briefly in Ref. \cite{ fund}). 

Among such consequences fixing the $\lambda_2$, $\lambda_5$, $\lambda_7$ ("antisymmetric") "gauge", one can mention fermionic rotary (axial) degrees of freedom ${\bf v}_1={\bf r}\times {\bf K}$ (with $\bf K$ being the polar colour vector, $ SU(2)_{\rm col}$ triplet).

Repeating the arguments  \cite{Pervush3},the source of the above fermionic (quark) rotary  degrees of freedom may be found in the interaction item
$ \sim   Z^a j_{Ia(0)}   $ in the Gauss-shell  reduced Hamiltonian of constraint-shell QCD.

This item implicates the zero-mode solution $Z^a$,  (\ref {zero}), to the Gauss law constraint (\ref {Gauss}) and the fermionic charge 
$$   j^{Ia(0)}\sim g \bar \psi^I (\lambda ^a/2)\gamma^0 \psi^I.
         $$
Here it would be set $a=2,5,7$ and $\gamma^0$ is the Dirac matrix; $\bar \psi^I$  and $\psi^I$  \cite{David3} are fermionic Dirac variable involving "small" Gribov multipliers $ v^{(0)}({\bf x})$.

Thus the existence of fermionic (quark) rotary  degrees of freedom  ${\bf v}_1$ (similar somewhat to rotary terms in molecules) can be associated lawfully with the "discrete vacuum geometry" assumed for Minkowskian constraint-shell QCD in the Dirac fundamental scheme \cite{Dir}.

\medskip
Secondly,  the  Mandelstam linearly increasing potential  $\sim Kr$ \cite{Cheng} in Minkowskian constraint-shell QCD "draws back in a background".

More exactly,  the Dirac fundamental scheme \cite{Dir} for Minkowskian constraint-shell QCD implicates a linear combination of well-known instantaneous ("four-fermionic") interaction potentials: Coulomb and "golden section" ones \cite{fund, David2, Pervush2,David3} (with coefficients depending somehow on the temperature $T$ and flavours mass scale $m$) as the source of hadronization and confinement in that model (see e.g. the papers \cite{Pervush2,Werner,Bogolubskaja}). 

Such linear combination may be got at finding the Green function of the Gauss law constraint equation in the BPS (Wu-Yang) monopole background \footnote{ Wu-Yang monopoles \cite{Wu}, as solutions to the equations of motions in the "purely YM" model, can be considered as spatial asymptotes for YM BPS monopoles.}.

\medskip
The important objection restricted   the role of the Mandelstam linearly increasing potential $\sim Kr$ in QCD was given also  in Ref. \cite{Leonidov}.

It turns out that for heavy quarconia, only the Coulomb potential  $\sim 1/r$ contributes in QCD.

Then the problem arises, to link  (continuously and smoothly) the Mandelstam linearly increasing  and Coulomb potentials in order to satisfy the heavy quarconia limit \cite{Leonidov}.

\medskip
As to the Mandelstam linearly increasing potential $\sim Kr$ \cite{Cheng}, the hypothesis can be suggested about including this potential at (finite) temperatures $T\neq 0$ in the linear combination with the Coulomb and "golden section" potentials. This can be done via a selection of the proper coefficient for the Mandelstam potential.

Among consequences for Minkowskian constraint-shell QCD of such including one can point out the "ordinary" annihilation mechanism \cite{Linde1} for monopole-antimonopole pairs via Abrikosov threads \cite{Al.S., Abrikosov} of the "magnetic" field (this mechanism was us discussed above). 

This is correlated closely with the dual-superconducting picture of confinement.

More exactly, at placing a monopole-antimonopole pair into the hot YM plasma, the typical  length of the Abrikosov thread ("magnetic" tube) between this pair becomes \cite{Linde, Linde1} $\Delta l\sim (g^2 T)^{-1}$. This implicates the temperature-depended Mandelstam potential in the shape $K(g^2 T)^{-1}$.

On the other hand, this including the temperature-depended Mandelstam potential in Minkowskian constraint-shell QCD can give rise to series of problems and questions. 

In particular, the problem  is to check satisfying the area law (as the confinement criterion) in the situation of specific "$\bf Z$-dominance" taking place actually in Minkowskian constraint-shell QCD and induced by $\bf Z$-vortices (thread topological defects) inside the appropriate vacuum manifold, assumed to be discrete and  given through (\ref{RYM}).  

\bigskip
 The insurmountable duality \cite{Al.S.} in specifying the sign of the vacuum "magnetic"  field $\bf B$ in the Bogomol'nyi equation (\ref{Bog}) taking place at assuming the discrete geometry (\ref{RYM}) for the vacuum manifold $R_{\rm YM}$ in the Minkowskian Higgs model with vacuum BPS monopoles quantized by Dirac was us discussed above. 

Remember herewith that the crucial point for this "insurmountable duality" is the existence of gauge transformations (\ref {dwaz}) \cite{Al.S.} (depending manifestly on thread solutions $A_\theta $ via Eq. (\ref {holonomija})).

Besides this duality, there is also the manifest symmetry of the "electric" action functional (\ref {rot}), squared by the topological momentum $P_N$ (given via Eq. (\ref {pin})), with respect to the changes in the sign of $P_N (k)$ ($k\in{\bf Z}$).

This becomes more obvious at recasting (\ref {rot}) to the look
\be
 \label{rotat}
 W_{N}= \int dt  \frac{ P_N^2 (t)}{2I}.
 \ee
Actually the said is equivalent  to    identifying  each two topologies with "opposite signs": $k$ and $-k$.\par
Thus, with taking account of the "discrete" factorization  (\ref{fact2}) for the residual $U(1)$ gauge symmetry, the "total" residual gauge symmetry group of the free rotator action (\ref{rot}) amounts  
\be \label{quot}
U_1=U_0\otimes {\bf Z}\otimes {\Bbb Z}_2.
\ee 
It is important to note here that this space is true manifold without singularities, i.e. does not create unnecessary problems. Since of the natural isomorphism $U_0\otimes {\bf Z}\cong U(1)\cong S^1$ (while $S^1$ is a manifold \footnote{Since $S^1$ is the set $\{x^2+y^2=1\}$, with $(x,y)\in R^2$ each point lies on a circle, and we can always find a small arc around that point that is homeomorphic to an open interval in 
$R$ (i.e., an open subset of 
$R$).
 This shows that $S^1$
 is locally 1-dimensional.

Also since $S^1\subset R^2$, it is a {\it Hausdorff space} as that inheriting the properties of $ R^2$.

The additional interesting way to see the manifold structure of $S^1$ is through an atlas. For instance, define two charts:

$\bullet$  Let $U_1=S^1/\{(0,1)\}$ and use stereographic projection from the point  $(0,1)$ onto the $x$-
axis. This gives a homeomorphism $\phi_1:U_1\to R$.

$\bullet$ Similarly,  Let $U_2=S^1/\{(0,-1)\}$ and use stereographic projection from  $(0,-1)$ onto the $x$-
axis, yielding $\phi_2:U_1\to R$.

These two charts cover $S^1$ and the transition maps between $\phi_1$ and $\phi_2$ (on their overlap $U_1\bigcap U_2$) are smooth (in fact, they are M$\ddot{\rm o}$bius transformations when written in the appropriate form). This atlas demonstrates that $S^1$ is not only a topological manifold but also a smooth 1-dimensional manifold.

}). 
It remains to show that the direct product $U_1=U_0\otimes {\bf Z}\otimes {\Bbb Z}_2$ is a manifold. But such a space is locally Euclidean because around any given point $(x, \pm 1)$ ($x\in U_0\otimes {\bf Z}$) we can take a neighborhood in $ U_0\otimes {\bf Z}$ homeomorphic to an open subset of $R^2$, and the fact that $\pm 1$ is isolated means that the product neighborhood is essentially the same open subset of $R^2$. 

\bigskip The same conclusion about identifying  each two topologies with "opposite signs" can be drawn about the vacuum "magnetic" energy \cite{David2}
\be 
\label{magn.e1}
\frac{1}{2}\int \limits_{\epsilon}^{\infty } d^3x [B_i ^a(\Phi_k)]^2 \equiv \frac{1}{2}V <B^2> =\frac 1{2\alpha_s}\int\limits_{\epsilon}^{\infty}\frac {dr}{r^2}\sim \frac 1 2  \frac 1
{\alpha_s\epsilon}= 2\pi \frac{gm}{g^2\sqrt{\lambda}}=\frac {2\pi} {g^2\epsilon}, 
\ee
squared by the vacuum "magnetic" tension $\bf B$, set, in turn, by the Bogomol'nyi equation (\ref{Bog}) and responsible for the superfluid properties of the BPS monopole  vacuum (suffered the Dirac fundamental quantization \cite{Dir}).

On the other hand, the Bogomol'nyi equation (\ref{Bog}) \cite{Gold}, associated the vacuum "magnetic" field $\bf B$ to the Higgs BPS monopole isomultiplet $\Phi^a$, specifies, indeed   to within a  sign, this vacuum "magnetic" field $\bf B$ and thus involves the invariance of the vacuum "magnetic" energy (\ref {magn.e1}) with respect to changes in signs of magnetic, that is equivalent of {\it topological},  charges.\par
The said implies  the modification of the "discrete" geometry  (\ref{RYM}) of the vacuum manifold $R_{\rm YM}$. \par 
Now it may be written down as 
\be \label{RYM1}
R'_{YM}= G_0/(U_0\otimes {\bf Z}\otimes {\mathbb Z}_2),
\ee
while  the "modified"  residual gauge symmetry group in the Minkowskian Higgs model with vacuum BPS monopoles quantized by Dirac is $U_1$, given by Eq. (\ref {quot}).
\par  
\par
It is easy to see (and we have comment this above) that the topological content of the Minkowskian Higgs model with vacuum BPS monopoles quantized by Dirac does not change at the modifications (\ref{quot}), (\ref{RYM1}).\par
Thus all the kinds of topological defects: thread and point hedgehog ones maintain in that model in spite of these modifications.

\end{document}